\documentclass[aps,superscriptaddress,nofootinbib,floatfix,notitlepage]{revtex4-2}
\usepackage[utf8x]{inputenc}
\pdfoutput=1
\usepackage{graphicx}
\usepackage{hyperref}
\usepackage{bm}
\usepackage{multirow}
\usepackage{array,booktabs,colortbl,multirow}
\usepackage{colortbl,xcolor,colordvi,color}
\usepackage{rotating}
\usepackage{placeins}

\newcommand{\cmrule}{\midrule[0.25mm]}

\newcolumntype{C}[1]{>{\centering\let\newline\\\arraybackslash\hspace{0pt}}m{#1}}

\usepackage[sort&compress]{natbib}

\begin{document}

\title{Global analysis of electroweak data in the Standard Model}

\author{J.~de Blas}
\affiliation{CAFPE and Departamento de F\'isica Te\'orica y del Cosmos, Universidad de Granada, Campus de Fuentenueva, E–18071 Granada, Spain}

\author{M.~Ciuchini}
\affiliation{INFN, Sezione di Roma Tre, Via della Vasca Navale 84, I-00146 Roma, Italy}

\author{E.~Franco}
\affiliation{INFN, Sezione di Roma, Piazzale A. Moro 2, I-00185 Roma, Italy}

\author{A.~Goncalves}
\affiliation{Physics Department, Florida State University,\\ Tallahassee, FL 32306-4350, USA}

\author{S.~Mishima}
\affiliation{Theory Center, IPNS, KEK, Tsukuba 305-0801, Japan}

\author{M.~Pierini}
\affiliation{CERN, 1211 Geneva 23, Switzerland}

\author{L.~Reina}
\affiliation{Physics Department, Florida State University,\\ Tallahassee, FL 32306-4350, USA}

\author{L.~Silvestrini}
\affiliation{INFN, Sezione di Roma, Piazzale A. Moro 2, I-00185 Roma, Italy}


\begin{abstract}
  We perform a global fit of electroweak data within the Standard Model, using state-of-the art
  experimental and theoretical results, including a determination of the electromagnetic coupling at the electroweak scale based on recent lattice calculations. In addition to the posteriors for all parameters and observables obtained from the global fit, we present indirect determinations for all parameters and
  predictions for all observables. 
  Furthermore, we present full predictions, obtained using only the experimental information on Standard Model parameters, and a fully indirect determination of Standard Model parameters using only experimental information on electroweak data. 
  Finally, we discuss in detail the compatibility of experimental data with the Standard Model and find a global \emph{p-value} of $0.5$.
\end{abstract}
\maketitle

In the Standard Model (SM) of ElectroWeak (EW) and strong
interactions, the $\mathrm{SU}(2)_L \otimes \mathrm{U}(1)_Y$ gauge
symmetry is hidden at low energies through the Higgs mechanism,
leaving only electromagnetism as a manifest symmetry. This hidden
symmetry endows the SM with calculable relations among masses and
couplings of EW bosons, and a huge theoretical effort in multi-loop
calculations has lead to the reduction of theoretical errors in these
relations down to $\mathcal{O}(10^{-4}-10^{-5})$~\cite{Sirlin:1980nh,Marciano:1980pb,Djouadi:1987gn,Djouadi:1987di,Kniehl:1989yc,Halzen:1990je,Kniehl:1991gu,Kniehl:1992dx,Barbieri:1992nz,Barbieri:1992dq,Djouadi:1993ss,Fleischer:1993ub,Fleischer:1994cb,Avdeev:1994db,Chetyrkin:1995ix,Chetyrkin:1995js,Degrassi:1996mg,Degrassi:1996ps,Degrassi:1999jd,Freitas:2000gg,vanderBij:2000cg,Freitas:2002ja,Awramik:2002wn,Onishchenko:2002ve,Awramik:2002vu,Awramik:2002wv,Awramik:2003ee,Awramik:2003rn,Faisst:2003px,Dubovyk:2016aqv,Dubovyk:2018rlg}. 
On the
experimental side, the monumental legacy of on- and off-peak
measurements of EW boson masses and couplings in $e^+ e^-$ collisions
at SLD, LEP, and LEP2 \cite{ALEPH:2005ab,LEP-2} has been supplemented by
TeVatron results \cite{ALEPH:2010aa,CDF:2016cry,CDF:2016vzt} and is being further improved
at the LHC \cite{Khachatryan:2015hba,Aaboud:2018zbu,Sirunyan:2018gqx,Sirunyan:2018goh,Sirunyan:2018mlv,ATLAS:2019ezb,Aaboud:2017svj,Aad:2015uau,ATLAS:2018gqq,Sirunyan:2018swq,Aaij:2015lka}. The discovery of the Higgs boson
\cite{Aad:2015zhl} completed the SM Lagrangian and marked a change of
perspective in the study of EW Precision Data (EWPD), allowing to
fully exploit their constraining power: all observables in the EW
sector can be precisely predicted, the overall consistency of the fit
can be assessed and SM parameters can be indirectly determined from
the fit. We carry out this program in a Bayesian framework, using
state-of-the-art experimental and theoretical results.

The study presented in this paper is carried out using the {\tt
  HEPfit} package~\cite{deBlas:2019okz,HEPfit}, a software tool to combine
direct and indirect contraints on the Standard Model and its
extensions.\footnote{The {\tt HEPfit} package is available under the
  GNU General Public License (GPL)~\cite{HEPfit}.} We perform a
Bayesian fit using the {\tt HEPfit} Markov Chain Monte Carlo (MCMC) engine
based on the BAT library~\cite{Caldwell:2008fw}.

\begin{table}[hp]
{
\begin{center}
\begin{tabular}{lccccc}
\toprule
\multicolumn{6}{c}{Global SM EW fit ({\it standard} scenario)} \\
&  &  &   &  &  \\ [-0.2cm]
& Measurement & Posterior & Prediction &1D Pull &nD Pull \\
\hline
$\alpha_{s}(M_{Z}^2)$ & $ 0.1177 \pm 0.0010 $ & $ 0.11792 \pm 0.00094 $  & $ 0.1198 \pm 0.0028 $& -0.7 &  \\ 
& & $[ 0.11606 , 0.11978 ]$ & $[ 0.1143 , 0.1253 ]$ & &  \\ 
$\Delta\alpha^{(5)}_{\mathrm{had}}(M_{Z}^2)$ & $ 0.02766 \pm 0.00010 $ & $ 0.027627 \pm 0.000096 $  & $ 0.02717 \pm 0.00037 $& 1.3 &  \\ 
& & $[ 0.027436 , 0.027815 ]$ & $[ 0.02646 , 0.02789 ]$ & &  \\ 
\rowcolor[gray]{.8} $M_Z$ [GeV] & $ 91.1875 \pm 0.0021 $ & $ 91.1883 \pm 0.0021 $  & $ 91.2047 \pm 0.0088 $& -1.9 &  \\ 
\rowcolor[gray]{.8} & & $[ 91.1842 , 91.1922 ]$ & $[ 91.1874 , 91.2217 ]$ & &  \\ 
$m_t$ [GeV] & $ 172.58 \pm 0.45 $ & $ 172.75 \pm 0.44 $  & $ 176.2 \pm 2.0 $& -1.8 &  \\ 
& & $[ 171.89 , 173.62 ]$ & $[ 172.2 , 180.0 ]$ & &  \\ 
$m_H$ [GeV] & $ 125.21 \pm 0.12 $ & $ 125.21 \pm 0.12 $  & $ 108.3 \pm 11.7 $& 1.3 &  \\ 
& & $[ 124.97 , 125.44 ]$ & $[ 90.1 , 137.4 ]$ & &  \\ 
\hline 
$M_W$ [GeV] & $ 80.379 \pm 0.012 $ & $ 80.3591 \pm 0.0052 $  & $ 80.3545 \pm 0.0057 $& 1.8 &  \\ 
& & $[ 80.3489 , 80.3692 ]$ & $[ 80.3433 , 80.3659 ]$ & &  \\ 
\hline 
$\Gamma_{W}$ [GeV] & $ 2.085 \pm 0.042 $ & $ 2.08827 \pm 0.00055 $  & $ 2.08829 \pm 0.00056 $& -0.1 &  \\ 
& & $[ 2.08719 , 2.08936 ]$ & $[ 2.08720 , 2.08938 ]$ & &  \\ 
\hline 
$\sin^2\theta_{\rm eff}^{\rm lept}(Q_{\rm FB}^{\rm had})$ & $ 0.2324 \pm 0.0012 $ & $ 0.231509 \pm 0.000056 $  & $ 0.231511 \pm 0.000058 $& 0.7 &  \\ 
& & $[ 0.231399 , 0.231619 ]$ & $[ 0.231398 , 0.231623 ]$ & &  \\ 
\hline 
$P_{\tau}^{\rm pol}=\mathcal{A}_\ell$ & $ 0.1465 \pm 0.0033 $ & $ 0.14712 \pm 0.00044 $  & $ 0.14713 \pm 0.00045 $& -0.2 &  \\ 
& & $[ 0.14625 , 0.14799 ]$ & $[ 0.14626 , 0.14801 ]$ & &  \\ 
\hline 
\rowcolor[gray]{.8} $\Gamma_{Z}$ [GeV] & $ 2.4955 \pm 0.0023 $ & $ 2.49443 \pm 0.00065 $  & $ 2.49423 \pm 0.00069 $& 0.5 &  \\ 
\rowcolor[gray]{.8} & & $[ 2.49317 , 2.49569 ]$ & $[ 2.49289 , 2.49558 ]$ & &  \\ 
\rowcolor[gray]{.8} $\sigma_{h}^{0}$ [nb] & $ 41.480 \pm 0.033 $ & $ 41.4908 \pm 0.0076 $  & $ 41.4927 \pm 0.0079 $& -0.4 &  \\ 
\rowcolor[gray]{.8} & & $[ 41.4756 , 41.5057 ]$ & $[ 41.4772 , 41.5086 ]$ & &  \\ 
\rowcolor[gray]{.8} $R^{0}_{\ell}$ & $ 20.767 \pm 0.025 $ & $ 20.7493 \pm 0.0080 $  & $ 20.7462 \pm 0.0087 $& 0.8 & 1.0 \\ 
\rowcolor[gray]{.8} & & $[ 20.7337 , 20.7651 ]$ & $[ 20.7291 , 20.7632 ]$ & &  \\ 
\rowcolor[gray]{.8} $A_{\rm FB}^{0, \ell}$ & $ 0.0171 \pm 0.0010 $ & $ 0.016234 \pm 0.000098 $  & $ 0.016225 \pm 0.000097 $& 0.9 &  \\ 
\rowcolor[gray]{.8} & & $[ 0.016043 , 0.016425 ]$ & $[ 0.016035 , 0.016417 ]$ & &  \\ 
\hline 
\rowcolor[gray]{.7} $\mathcal{A}_{\ell}$ (SLD) & $ 0.1513 \pm 0.0021 $ & $ 0.14712 \pm 0.00044 $  & $ 0.14713 \pm 0.00046 $& 1.9 &  \\ 
\rowcolor[gray]{.7} & & $[ 0.14625 , 0.14799 ]$ & $[ 0.14622 , 0.14803 ]$ & &  \\ 
\rowcolor[gray]{.7} $R^{0}_{b}$ & $ 0.21629 \pm 0.00066 $ & $ 0.215878 \pm 0.000100 $  & $ 0.21587 \pm 0.00010 $& 0.6 &  \\ 
\rowcolor[gray]{.7} & & $[ 0.215681 , 0.216075 ]$ & $[ 0.21567 , 0.21607 ]$ & &  \\ 
\rowcolor[gray]{.7} $R^{0}_{c}$ & $ 0.1721 \pm 0.0030 $ & $ 0.172205 \pm 0.000054 $  & $ 0.172206 \pm 0.000053 $& 0.0 &  \\ 
\rowcolor[gray]{.7} & & $[ 0.172100 , 0.172310 ]$ & $[ 0.172101 , 0.172311 ]$ & &  \\ 
\rowcolor[gray]{.7} $A_{\rm FB}^{0, b}$ & $ 0.0996 \pm 0.0016 $ & $ 0.10314 \pm 0.00031 $  & $ 0.10315 \pm 0.00033 $& -2.2 & 1.3 \\ 
\rowcolor[gray]{.7} & & $[ 0.10253 , 0.10375 ]$ & $[ 0.10251 , 0.10378 ]$ & &  \\ 
\rowcolor[gray]{.7} $A_{\rm FB}^{0, c}$ & $ 0.0707 \pm 0.0035 $ & $ 0.07369 \pm 0.00023 $  & $ 0.07370 \pm 0.00024 $& -0.9 &  \\ 
\rowcolor[gray]{.7} & & $[ 0.07324 , 0.07414 ]$ & $[ 0.07322 , 0.07416 ]$ & &  \\ 
\rowcolor[gray]{.7} $\mathcal{A}_b$ & $ 0.923 \pm 0.020 $ & $ 0.934738 \pm 0.000040 $  & $ 0.934739 \pm 0.000040 $& -0.6 &  \\ 
\rowcolor[gray]{.7} & & $[ 0.934661 , 0.934817 ]$ & $[ 0.934661 , 0.934818 ]$ & &  \\ 
\rowcolor[gray]{.7} $\mathcal{A}_c$ & $ 0.670 \pm 0.027 $ & $ 0.66782 \pm 0.00022 $  & $ 0.66783 \pm 0.00022 $& 0.0 &  \\ 
\rowcolor[gray]{.7} & & $[ 0.66739 , 0.66825 ]$ & $[ 0.66740 , 0.66825 ]$ & &  \\ 
\hline 
$\mathcal{A}_s$ & $ 0.895 \pm 0.091 $ & $ 0.935651 \pm 0.000040 $  & $ 0.935651 \pm 0.000040 $& -0.4 &  \\ 
& & $[ 0.935573 , 0.935730 ]$ & $[ 0.935572 , 0.935729 ]$ & &  \\ 
BR$_{W\to\ell\bar\nu_\ell}$ & $ 0.10860 \pm 0.00090 $ & $ 0.108381 \pm 0.000022 $  & $ 0.108380 \pm 0.000022 $& 0.2 &  \\ 
& & $[ 0.108337 , 0.108424 ]$ & $[ 0.108337 , 0.108423 ]$ & &  \\ 
$\sin^2\theta_{\rm eff}^{\mathrm{lept}}$ (HC) & $ 0.23143 \pm 0.00025 $ & $ 0.231509 \pm 0.000056 $  & $ 0.231511 \pm 0.000058 $& -0.3 &  \\ 
& & $[ 0.231399 , 0.231619 ]$ & $[ 0.231398 , 0.231623 ]$ & &  \\ 
\hline 
$R_{uc}$ & $ 0.1660 \pm 0.0090 $ & $ 0.172227 \pm 0.000032 $  & $ 0.172228 \pm 0.000032 $& -0.7 &  \\ 
& & $[ 0.172166 , 0.172290 ]$ & $[ 0.172166 , 0.172290 ]$ & &  \\ 
\bottomrule
\end{tabular}
\end{center}
}
\caption{
  Experimental measurement, result of the global fit, prediction, and pull for the five input
  parameters ($\alpha_s(M_Z^2)$, $\Delta \alpha^{(5)}_{\mathrm{had}}(M_Z^2)$, $M_Z$,
  $m_t$, $m_H$), and for the set of EWPO considered in the fit, in the \emph{standard} scenario for $m_t$ and $m_H$. For the results of the global 
  fit and for the predictions, the $95\%$ probability range is reported in square brackets. The values in
  the column \emph{Prediction} are determined without using the
  corresponding experimental information. Pulls are calculated
  both as individual pulls (\emph{1D Pull}) and as global pulls
  (\emph{nD Pull}) for sets of correlated observables, and are given in units of standard
  deviations. Groups of correlated observables
  are identified by shades of grey.} 
\label{tab:SMfit-standard}
\end{table}

\begin{table}[hp]
{
\begin{center}
\begin{tabular}{lccccc}
\toprule
\multicolumn{6}{c}{Global SM EW fit ({\it conservative} scenario)} \\
&  &  &   &  &  \\ [-0.2cm]
& Measurement & Posterior & Prediction &1D Pull &nD Pull \\
\hline 
$\alpha_{s}(M_{Z}^2)$ & $ 0.1177 \pm 0.0010 $ & $ 0.11793 \pm 0.00094 $  & $ 0.1199 \pm 0.0028 $& -0.7 &  \\ 
& & $[ 0.11610 , 0.11979 ]$ & $[ 0.1143 , 0.1254 ]$ & &  \\ 
$\Delta\alpha^{(5)}_{\mathrm{had}}(M_{Z}^2)$ & $ 0.02766 \pm 0.00010 $ & $ 0.027631 \pm 0.000097 $  & $ 0.02721 \pm 0.00039 $& 1.1 &  \\ 
& & $[ 0.027441 , 0.027823 ]$ & $[ 0.02646 , 0.02797 ]$ & &  \\ 
\rowcolor[gray]{.8} $M_Z$ [GeV] & $ 91.1875 \pm 0.0021 $ & $ 91.1881 \pm 0.0021 $  & $ 91.2045 \pm 0.0094 $& -1.8 &  \\ 
\rowcolor[gray]{.8} & & $[ 91.1841 , 91.1922 ]$ & $[ 91.1859 , 91.2231 ]$ & &  \\ 
$m_t$ [GeV] & $ 172.6 \pm 1.0 $ & $ 173.31 \pm 0.90 $  & $ 176.1 \pm 2.0 $& -1.6 &  \\ 
& & $[ 171.55 , 175.06 ]$ & $[ 172.2 , 180.0 ]$ & &  \\ 
$m_H$ [GeV] & $ 125.21 \pm 0.21 $ & $ 125.21 \pm 0.21 $  & $ 109.7 \pm 12.6 $& 1.2 &  \\ 
& & $[ 124.80 , 125.62 ]$ & $[ 89.9 , 141.2 ]$ & &  \\ 
\hline 
$M_W$ [GeV] & $ 80.379 \pm 0.012 $ & $ 80.3619 \pm 0.0064 $  & $ 80.3549 \pm 0.0077 $& 1.7 &  \\ 
& & $[ 80.3491 , 80.3746 ]$ & $[ 80.3398 , 80.3700 ]$ & &  \\ 
\hline 
$\Gamma_{W}$ [GeV] & $ 2.085 \pm 0.042 $ & $ 2.08850 \pm 0.00063 $  & $ 2.08849 \pm 0.00063 $& -0.1 &  \\ 
& & $[ 2.08725 , 2.08974 ]$ & $[ 2.08726 , 2.08975 ]$ & &  \\ 
\hline 
$\sin^2\theta_{\rm eff}^{\rm lept}(Q_{\rm FB}^{\rm had})$ & $ 0.2324 \pm 0.0012 $ & $ 0.231497 \pm 0.000058 $  & $ 0.231498 \pm 0.000060 $& 0.8 &  \\ 
& & $[ 0.231382 , 0.231612 ]$ & $[ 0.231380 , 0.231617 ]$ & &  \\ 
\hline 
$P_{\tau}^{\rm pol}=\mathcal{A}_\ell$ & $ 0.1465 \pm 0.0033 $ & $ 0.14722 \pm 0.00046 $  & $ 0.14724 \pm 0.00046 $& -0.2 &  \\ 
& & $[ 0.14632 , 0.14813 ]$ & $[ 0.14632 , 0.14814 ]$ & &  \\ 
\hline 
\rowcolor[gray]{.8} $\Gamma_{Z}$ [GeV] & $ 2.4955 \pm 0.0023 $ & $ 2.49454 \pm 0.00066 $  & $ 2.49434 \pm 0.00070 $& 0.5 &  \\ 
\rowcolor[gray]{.8} & & $[ 2.49323 , 2.49584 ]$ & $[ 2.49297 , 2.49573 ]$ & &  \\ 
\rowcolor[gray]{.8} $\sigma_{h}^{0}$ [nb] & $ 41.480 \pm 0.033 $ & $ 41.4912 \pm 0.0077 $  & $ 41.4931 \pm 0.0080 $& -0.4 &  \\ 
\rowcolor[gray]{.8} & & $[ 41.4761 , 41.5062 ]$ & $[ 41.4774 , 41.5091 ]$ & &  \\ 
\rowcolor[gray]{.8} $R^{0}_{\ell}$ & $ 20.767 \pm 0.025 $ & $ 20.7492 \pm 0.0080 $  & $ 20.7458 \pm 0.0087 $& 0.8 & 0.9 \\ 
\rowcolor[gray]{.8} & & $[ 20.7335 , 20.7650 ]$ & $[ 20.7290 , 20.7629 ]$ & &  \\ 
\rowcolor[gray]{.8} $A_{\rm FB}^{0, \ell}$ & $ 0.0171 \pm 0.0010 $ & $ 0.01626 \pm 0.00010 $  & $ 0.01625 \pm 0.00010 $& 0.9 &  \\ 
\rowcolor[gray]{.8} & & $[ 0.01606 , 0.01645 ]$ & $[ 0.01605 , 0.01645 ]$ & &  \\ 
\hline 
\rowcolor[gray]{.7} $\mathcal{A}_{\ell}$ (SLD) & $ 0.1513 \pm 0.0021 $ & $ 0.14722 \pm 0.00046 $  & $ 0.14724 \pm 0.00048 $& 1.9 &  \\ 
\rowcolor[gray]{.7} & & $[ 0.14632 , 0.14813 ]$ & $[ 0.14629 , 0.14819 ]$ & &  \\ 
\rowcolor[gray]{.7} $R^{0}_{b}$ & $ 0.21629 \pm 0.00066 $ & $ 0.21586 \pm 0.00010 $  & $ 0.21585 \pm 0.00010 $& 0.7 &  \\ 
\rowcolor[gray]{.7} & & $[ 0.21565 , 0.21606 ]$ & $[ 0.21564 , 0.21606 ]$ & &  \\ 
\rowcolor[gray]{.7} $R^{0}_{c}$ & $ 0.1721 \pm 0.0030 $ & $ 0.172212 \pm 0.000055 $  & $ 0.172212 \pm 0.000054 $& 0.0 &  \\ 
\rowcolor[gray]{.7} & & $[ 0.172106 , 0.172318 ]$ & $[ 0.172106 , 0.172319 ]$ & &  \\ 
\rowcolor[gray]{.7} $A_{\rm FB}^{0, b}$ & $ 0.0996 \pm 0.0016 $ & $ 0.10321 \pm 0.00033 $  & $ 0.10323 \pm 0.00034 $& -2.2 & 1.3 \\ 
\rowcolor[gray]{.7} & & $[ 0.10257 , 0.10384 ]$ & $[ 0.10255 , 0.10389 ]$ & &  \\ 
\rowcolor[gray]{.7} $A_{\rm FB}^{0, c}$ & $ 0.0707 \pm 0.0035 $ & $ 0.07374 \pm 0.00024 $  & $ 0.07375 \pm 0.00025 $& -0.9 &  \\ 
\rowcolor[gray]{.7} & & $[ 0.07326 , 0.07422 ]$ & $[ 0.07325 , 0.07425 ]$ & &  \\ 
\rowcolor[gray]{.7} $\mathcal{A}_b$ & $ 0.923 \pm 0.020 $ & $ 0.934741 \pm 0.000040 $  & $ 0.934741 \pm 0.000040 $& -0.6 &  \\ 
\rowcolor[gray]{.7} & & $[ 0.934662 , 0.934819 ]$ & $[ 0.934662 , 0.934820 ]$ & &  \\ 
\rowcolor[gray]{.7} $\mathcal{A}_c$ & $ 0.670 \pm 0.027 $ & $ 0.66787 \pm 0.00023 $  & $ 0.66788 \pm 0.00023 $& 0.1 &  \\ 
\rowcolor[gray]{.7} & & $[ 0.66742 , 0.66832 ]$ & $[ 0.66742 , 0.66833 ]$ & &  \\ 
\hline 
$\mathcal{A}_s$ & $ 0.895 \pm 0.091 $ & $ 0.935660 \pm 0.000042 $  & $ 0.935659 \pm 0.000042 $& -0.4 &  \\ 
& & $[ 0.935577 , 0.935743 ]$ & $[ 0.935578 , 0.935742 ]$ & &  \\ 
BR$_{W\to\ell\bar\nu_\ell}$ & $ 0.10860 \pm 0.00090 $ & $ 0.108380 \pm 0.000022 $  & $ 0.108380 \pm 0.000022 $& 0.2 &  \\ 
& & $[ 0.108337 , 0.108424 ]$ & $[ 0.108337 , 0.108424 ]$ & &  \\ 
$\sin^2\theta_{\rm eff}^{\mathrm{lept}}$ (HC) & $ 0.23143 \pm 0.00025 $ & $ 0.231497 \pm 0.000058 $  & $ 0.231498 \pm 0.000060 $& -0.3 &  \\ 
& & $[ 0.231382 , 0.231612 ]$ & $[ 0.231380 , 0.231617 ]$ & &  \\ 
\hline 
$R_{uc}$ & $ 0.1660 \pm 0.0090 $ & $ 0.172234 \pm 0.000033 $  & $ 0.172234 \pm 0.000033 $& -0.7 &  \\ 
& & $[ 0.172170 , 0.172299 ]$ & $[ 0.172170 , 0.172299 ]$ & &  \\ 
\bottomrule
\end{tabular}
\end{center}
}
\caption{
  Same as Table \ref{tab:SMfit-standard} in the \emph{conservative} scenario for the errors on $m_t$ and $m_H$.} 
\label{tab:SMfit-conservative}
\end{table}

The list of parameters and EW precision 
observables (EWPO) included in our study is shown in
Tables~\ref{tab:SMfit-standard} and \ref{tab:SMfit-conservative},
where we present results for the SM fit of EWPD
both in a \textit{standard} (Table~\ref{tab:SMfit-standard}) and in a \textit{conservative} (Table~\ref{tab:SMfit-conservative})
scenario, depending on the assumptions made in combining different
measurements of $m_t$ and $m_H$, as
described below.

The main framework of the EW fit as implemented in {\tt HEPfit} can be
found in Refs.~\cite{Ciuchini:2013pca,deBlas:2016ojx}, to which we refer
the reader for a more detailed description of how various EWPO have been
implemented, and for a complete account of the
literature on which such implementations are based. With respect to
our previous studies, we now also take into account the latest
developments on the theory side, such as the recent calculation of the
2-loop EW bosonic corrections to $\sin^2\theta_{\mathrm{eff}}^{b}$
\cite{Dubovyk:2016aqv}, as well as the full 2-loop corrections to the
partial decays of the $Z$ from Ref.~\cite{Dubovyk:2018rlg}. As explained in
Ref.~\cite{Dubovyk:2018rlg}, such corrections are very small and indeed we find
they have no noticeable effect on the fit results.\footnote{The recent evaluation of the leading fermionic three-loop corrections to EWPO~\cite{Chen:2020xzx,Chen:2020xot} results in even smaller effects, which have been neglected in our fits.}

Among the input parameters, $G_\mu$ and $\alpha(0)$ are fixed
($G_\mu=1.1663787\times10^{-5}$~GeV$^{-2}$ and
$\alpha(0)=1/137.035999139$~\cite{Tanabashi:2018oca}), while
$\alpha_s(M_Z^2)$, $\Delta\alpha^{(5)}_{\mathrm{had}}(M_Z^2)$, $M_Z$,
$m_t$, and $m_H$ are varied in the MCMC process.  Compared to our latest
analysis~\cite{deBlas:2016ojx}, the EW fit presented in this paper
contains several updates that are summarized below:\footnote{For recent comprehensive reviews of both theoretical and
  experimental inputs to EW precision fits see
 Ref.~\cite{Tanabashi:2018oca} and Ref.~\cite{Erler:2019hds}.}
\begin{enumerate}
\item The value of the strong coupling constant at the $Z$ pole has
    been updated with the last average from the Particle Data Group
    (PDG), $\alpha_s (M_Z^2)=0.1179\pm 0.0010$~\cite{Zyla:2020zbs}.
    To avoid double counting the experimental information on EW observables, we exclude from this combination the
     indirect determination from the EW fit
    and obtain
    $\alpha_s (M_Z^2)=0.1177\pm 0.0010$.
    This updates the previous average used in
   Ref.~\cite{deBlas:2016ojx}, $\alpha_s (M_Z^2)=0.1179 \pm 0.0012$.
   The small differences, both in the central value and error,
   have only a small effect on the global EW fit.
\item The value for the five-flavour hadronic contribution to the QED
coupling constant at the $Z$-boson mass has also been recently updated
by several groups, with mostly compatible results (see Ref.~\cite{Erler:2019hds} for a
more comprehensive discussion).  In our study we make use of the
lattice determinations of the euclidean correlation function $\hat
\Pi_{n_f=4}(-4 \mathrm{GeV}^2) = 0.0712 \pm 0.0002$ from
Ref.~\cite{Borsanyi:2017zdw} (Table S3) and of the bottom
quark contribution $\hat \Pi_b(-4\mathrm{GeV}^2) = 0.00013$ from
Ref.~\cite{Colquhoun:2014ica}, 
which combined give $$\Delta
\alpha_{\mathrm{had}}^{(5)}(-4\mathrm{GeV}^2)=4 \pi \alpha 
\left(
  \hat
\Pi_{n_f=4}(-4 \mathrm{GeV}^2) + \hat \Pi_b(-4\mathrm{GeV}^2)
\right)=0.00654 \pm
0.00002.$$ Running perturbatively to the scale $-M_Z^2$ and continuing
analytically to Minkowski spacetime according to
Ref.~\cite{Blondel:2019vdq} leads to  $\Delta
\alpha_{\mathrm{had}}^{(5)}(M_Z^2)=0.02766 \pm 0.00010$,
compatible with, but more precise than, the value
we previously used~\cite{deBlas:2016ojx}, $\Delta
\alpha_{\mathrm{had}}^{(5)}(M_Z^2)=0.02750 \pm 0.00033$.
\item For the top-quark mass, Ref.~\cite{deBlas:2016ojx} used the 2014
  world average from ATLAS and the Tevatron experiments. Since then
  several updated measurements have become available, with individual
  uncertainties exceeding that of the 2014 average. We have
  therefore reconsidered and updated the value of and uncertainty on
  $m_t$ that is used on the current EW precision fit. In this study we
  consider: i) the 2016
  Tevatron~\cite{TevatronElectroweakWorkingGroup:2016lid} combination;
  ii) the 2015
  CMS Run 1 combination~\cite{Khachatryan:2015hba};
  iii) the combination of ATLAS Run 1 results in Ref.~\cite{Aaboud:2018zbu};
  iii) the CMS Run 2 measurements in the dilepton, lepton+jets and all-jet channels \cite{Sirunyan:2018gqx,Sirunyan:2018goh,Sirunyan:2018mlv};
  and iv) the ATLAS Run 2 result from the lepton+jet channel~\cite{ATLAS:2019ezb}. 
    Unfortunately, combining these
  different measurements is non-trivial due to the correlations
  between theoretical errors and several of the systematic uncertainties of the
  different measurements. For the purpose of this paper, we consider
  a correlated combination between the different measurements,\footnote{We do not consider the CMS measurement using the single-top
  channel~\cite{Sirunyan:2017huu}, but we checked it has a negligible impact on the average.} assuming
  the linear correlation coefficient between two systematic uncertainties to be written as $\rho_{ij}^\mathrm{sys}=\mathrm{min}\left\{\sigma_i^\mathrm{sys}, \sigma_j^\mathrm{sys}\right\}/\mathrm{max}\left\{\sigma_i^\mathrm{sys}, \sigma_j^\mathrm{sys}\right\}$.
  This results in $m_t=172.58\pm 0.45$ GeV.
  In performing this combination, we note that, while the LHC measurements of $m_t$
  are reasonably consistent with each other, there is some tension
  between the ATLAS and CMS lepton+jet values. This is also the case
  between the LHC and Tevatron $m_t$ combinations, and while these tensions could be just
  due to statistical fluctuations, in the worst case scenario they
  could indicate that some of the systematics included in these measurements
  have been underestimated. A common way of dealing with this issue is to use a
  rescaled error following the PDG average method~\cite{Zyla:2020zbs}. In our case
  the resulting uncertainty would turn out to be unreasonably large, $\sim 1.7$ GeV.
  For the purpose of this paper we illustrate the impact of the top-mass
  uncertainty on the SM precision fits by considering two scenarios:
  one where we use the \textit{standard} error of
  $\delta m_t=0.45$~GeV and one where we consider a more
  \textit{conservative} error of $\delta m_t=1$~GeV.
  As we will see, the two
  different scenarios lead at the moment to very similar results
  since the parametric uncertainties are subleading with respect to
  the experimental ones. 
\item The Higgs-boson mass, whose value after Run 1 was
  $m_H=125.09\pm 0.24$ GeV~\cite{Aad:2015zhl} (as used in
  Ref.~\cite{deBlas:2016ojx}), has now also been measured in Run 2 both
  by ATLAS~\cite{Aaboud:2018wps,ATLAS:2020coj} and
  CMS~\cite{Sirunyan:2017exp,Sirunyan:2020xwk} in the $4\ell$ and
  $\gamma\gamma$ channels. 
  We follow the same combination procedure as for $m_t$ and combine
  all the different measurements to obtain $m_H=125.21\pm 0.12$ GeV as the {\it standard}
  input value for our fits. Some tension between the ATLAS and CMS Run 2 combinations
  is also present in this case. 
  This tension does not have a visible impact in the fits, since the parametric uncertainty associated to $m_H$ is negligible for $\delta m_H$ up to O(10 GeV). Nevertheless, we consider an
  uncertainty $\delta m_H=0.21$ GeV for the {\it conservative} scenario, obtained using the
  PDG scaling method~\cite{Zyla:2020zbs}. 
\item The ATLAS collaboration presented their first direct
  determination of the $W$-boson mass, $M_W=80.370\pm 0.019$ GeV
  \cite{Aaboud:2017svj}, with an uncertainty comparable to the current
  LEP2+Tevatron average. Assuming
  the absence of significant correlations between the LHC and Tevatron
  determinations, one can compute an approximate ``world average" of
  $M_W=80.379\pm0.012$ GeV.  Although different from the LEP2+Tevatron
  world average used in Ref.~\cite{deBlas:2016ojx},
  $M_W=80.385\pm0.015$ GeV, the update of $M_W$ has a very minor
  effect on the SM fits.\footnote{The very recent result on $M_W$ from the LHCb Collaboration \cite{LHCb:2021bjt} will be included in future updates of our fit.}
  \item The determination of the effective leptonic weak mixing angle,
   $\sin^2\theta_{\mathrm{eff}}^{\mathrm{lept}}$, at hadron colliders has
    also been updated. Using the same procedure as for the Higgs-boson and
    top-quark masses, we combine the ATLAS~\cite{Aad:2015uau,ATLAS:2018gqq} and
    CMS~\cite{Sirunyan:2018swq} measurements, the Tevatron
    determinations in Ref.~\cite{Aaltonen:2018dxj}, and the LHCb
    measurement in Ref.~\cite{Aaij:2015lka}.
    This combination is done separately for the measurements in the electron and muon channels, yielding
    $\sin^2\theta_{\mathrm{eff}}^{ee}=0.23175\pm 0.00029$ and $\sin^2\theta_{\mathrm{eff}}^{\mu\mu}=0.23093\pm 0.00039$, respectively. We also obtain a combination assuming lepton universality, $\sin^2\theta_{\mathrm{eff}}^{\mathrm{lept}}=0.23143\pm 0.00025$.
    In this last case, we note that there is some tension between the CDF and D0 values, but this would only
    result in a small rescaling of the error and it is ignored here. 
  \item The updates in the $Z$-lineshape observables reported in Ref. \cite{Janot:2019oyi} have been included.
  Compared to Ref.~\cite{ALEPH:2005ab}, these updates are due to the use of more accurate calculations of the Bhabha cross section, which lead to a better understanding of any systematic bias on the integrated luminosity. Only the $Z$ width, $\Gamma_Z$, the  hadronic cross section at the $Z$ peak, $\sigma_h^0$, and its correlations with other $Z$-lineshape observables are noticeably affected by these updates.
 \item We have included the update in the determination of the forward-backward asymmetry of the bottom quark, $A_{FB}^{0,b}$, after
 taking into account the massive $O(\alpha_s^2)$ corrections in $e^+ e^- \to b\bar{b}$ at the $Z$ pole~\cite{Bernreuther:2016ccf}.
 As we will see, these corrections slightly reduce the longstanding tension between the experimental measurement of this observable and its SM prediction.
\end{enumerate}

Apart from these updates, we have also extended the EW fit by including the following extra observables:
\begin{enumerate}
\setcounter{enumi}{8}
{\item The determination of the $s$-quark asymmetry parameter ${\cal A}_s$ at SLD~\cite{Abe:2000uc}.}
{\item The PDG average of the different LEP experiment determinations of the ratio $R_{uc}\equiv \frac 12 \Gamma_{Z\to u\bar{u} + c\bar{c}}/\Gamma_{Z\to \mathrm{had}}$~\cite{Zyla:2020zbs}.}
{\item The leptonic branching ratio of the $W$ boson, ${\rm BR}_{W\to \ell \bar\nu_\ell}\equiv \Gamma_{W^- \to \ell^- \bar\nu_\ell}/\Gamma_W$ ~\cite{Zyla:2020zbs}.}
\end{enumerate}
\begin{table}[tb]
\centering
\begin{tabular}{ll|lll|ll|ll}
  \toprule
  & & & & &\multicolumn{2}{c|}{\emph{standard} scenario} & \multicolumn{2}{c}{\emph{conservative} scenario}\\
&  Prediction &
\quad ${\alpha_s}(M_{Z}^2)$ &
\ ${\Delta\alpha_{\rm had}^{(5)}}(M_{Z}^2)$ &
\quad ${M_Z}$ &
\quad ${m_t}$ & Total & \quad ${m_t}$ & Total
\\
  \cmrule
$M_W$ [GeV] & $ 80.3545 $
& $\pm 0.0006 $ 
& $\pm 0.0018 $ 
& $\pm 0.0027 $ 
& $\pm 0.0027 $ 
& $\pm 0.0042 $ 
& $\pm 0.0060 $ 
& $\pm 0.0069 $ 
\\ 
$\Gamma_{W}$ [GeV] & $ 2.08782 $
& $\pm 0.00040 $ 
& $\pm 0.00014 $ 
& $\pm 0.00021 $ 
& $\pm 0.00021 $ 
& $\pm 0.00052 $ 
& $\pm 0.00047 $ 
& $\pm 0.00066 $ 
\\ 
$\Gamma_{Z}$ [GeV] & $ 2.49414 $
& $\pm 0.00049 $ 
& $\pm 0.00010 $ 
& $\pm 0.00021 $ 
& $\pm 0.00010 $ 
& $\pm 0.00056 $ 
& $\pm 0.00023 $ 
& $\pm 0.00060 $ 
\\ 
$\sigma_{h}^{0}$ [nb] & $ 41.4929 $
& $\pm 0.0049 $ 
& $\pm 0.0001 $ 
& $\pm 0.0020 $ 
& $\pm 0.0003 $ 
& $\pm 0.0053 $ 
& $\pm 0.0007 $ 
& $\pm 0.0053 $ 
\\ 
$\sin^2\theta_{\rm eff}^{\rm lept}$ & $ 0.231534 $
& $\pm 0.000003 $ 
& $\pm 0.000035 $ 
& $\pm 0.000015 $ 
& $\pm 0.000013 $ 
& $\pm 0.000041 $ 
& $\pm 0.000030 $ 
& $\pm 0.000048 $ 
\\ 
$\mathcal{A}_{\ell}$ & $ 0.14692 $
& $\pm 0.00003 $ 
& $\pm 0.00028 $ 
& $\pm 0.00012 $ 
& $\pm 0.00010 $ 
& $\pm 0.00032 $ 
& $\pm 0.00023 $ 
& $\pm 0.00038 $ 
\\ 
$\mathcal{A}_c$ & $ 0.66775 $
& $\pm 0.00001 $ 
& $\pm 0.00012 $ 
& $\pm 0.00005 $ 
& $\pm 0.00005 $ 
& $\pm 0.00014 $ 
& $\pm 0.00011 $ 
& $\pm 0.00017 $ 
\\ 
$\mathcal{A}_b$ & $ 0.934727 $
& $\pm 0.000001 $ 
& $\pm 0.000023 $ 
& $\pm 0.000010 $ 
& $\pm 0.000003 $ 
& $\pm 0.000025 $ 
& $\pm 0.000007 $ 
& $\pm 0.000026 $ 
\\ 
$A_{\rm FB}^{0, \ell}$ & $ 0.016191 $
& $\pm 0.000006 $ 
& $\pm 0.000060 $ 
& $\pm 0.000026 $ 
& $\pm 0.000023 $ 
& $\pm 0.000070 $ 
& $\pm 0.000052 $ 
& $\pm 0.000084 $ 
\\ 
$A_{\rm FB}^{0, c}$ & $ 0.07358 $
& $\pm 0.00001 $ 
& $\pm 0.00015 $ 
& $\pm 0.00006 $ 
& $\pm 0.00006 $ 
& $\pm 0.00018 $ 
& $\pm 0.00013 $ 
& $\pm 0.00021 $ 
\\ 
$A_{\rm FB}^{0, b}$ & $ 0.10300 $
& $\pm 0.00002 $ 
& $\pm 0.00020 $ 
& $\pm 0.00008 $ 
& $\pm 0.00007 $ 
& $\pm 0.00023 $ 
& $\pm 0.00016 $ 
& $\pm 0.00027 $ 
\\ 
$R^{0}_{\ell}$ & $ 20.7464 $
& $\pm 0.0062 $ 
& $\pm 0.0006 $ 
& $\pm 0.0003 $ 
& $\pm 0.0002 $ 
& $\pm 0.0063 $ 
& $\pm 0.0004 $ 
& $\pm 0.0063 $ 
\\ 
$R^{0}_{c}$ & $ 0.172198 $
& $\pm 0.000020 $ 
& $\pm 0.000002 $ 
& $\pm 0.000001 $ 
& $\pm 0.000005 $ 
& $\pm 0.000020 $ 
& $\pm 0.000011 $ 
& $\pm 0.000023 $ 
\\ 
$R^{0}_{b}$ & $ 0.215880 $
& $\pm 0.000011 $ 
& $\pm 0.000001 $ 
& $\pm 0.000000 $ 
& $\pm 0.000015 $ 
& $\pm 0.000019 $ 
& $\pm 0.000034 $ 
& $\pm 0.000035 $ 
\\ 
BR$_{W\to\ell\bar\nu_\ell}$ & $ 0.108386 $
& $\pm 0.000024 $ 
& $\pm 0.000000 $ 
& $\pm 0.000000 $ 
& $\pm 0.000000 $ 
& $\pm 0.000024 $ 
& $\pm 0.000000 $ 
& $\pm 0.000024 $ 
\\ 
$\mathcal{A}_{s}$ & $ 0.935637 $
& $\pm 0.000002 $ 
& $\pm 0.000022 $ 
& $\pm 0.000010 $ 
& $\pm 0.000009 $ 
& $\pm 0.000026 $ 
& $\pm 0.000020 $ 
& $\pm 0.000031 $ 
\\ 
$R_{uc}$ & $ 0.172220 $
& $\pm 0.000019 $ 
& $\pm 0.000002 $ 
& $\pm 0.000001 $ 
& $\pm 0.000005 $ 
& $\pm 0.000020 $ 
& $\pm 0.000011 $ 
& $\pm 0.000023 $ 
\\ 
\bottomrule
\end{tabular}
\caption{SM predictions computed using the theoretical expressions for
  the EWPO without the corresponding experimental constraints, and
  individual uncertainties associated with each
  input parameter, except for $m_H$ (see text).}
\label{tab:SMpred}
\end{table}
We use flat priors for
all the SM input parameters, and include the information of their
experimental measurements in the likelihood. We assume that all
experimental distributions are Gaussian. 
The known {\it intrinsic} theoretical uncertainties due to missing
higher-order corrections to EWPO are also included in the fits, using the results of Ref.~\cite{Dubovyk:2018rlg} to which we refer for more
details. 
The main theory uncertainties we consider are:
\begin{eqnarray}
    &&\delta_\mathrm{th} M_W = 4\, \mathrm{MeV}\,,\quad \delta_\mathrm{th} \sin^2{\theta_W} = 5\cdot 10^{-5}\,,\quad \delta_\mathrm{th} \Gamma_{Z} = 0.4\, \mathrm{MeV} \,,\quad
    \delta_\mathrm{th} \sigma^0_\mathrm{had} = 6\, \mathrm{pb} \,,\quad \nonumber \\
    &&\delta_\mathrm{th} R^0_\ell = 0.006 \,,\quad \delta_\mathrm{th} R^0_c = 0.00005 \,,\quad
    \delta_\mathrm{th} R^0_b = 0.0001 \,.\quad
    \label{eq:therr}
\end{eqnarray}
These uncertainties are implemented in
the fit as nuisance parameters with Gaussian prior distributions.  Theoretical uncertainties are
still small compared to the experimental ones and, therefore, they have a very
small impact on the fit. The same applies to the {\it parametric}
theory uncertainties, obtained by propagating the experimental errors
of the SM inputs into the predictions for the EWPO.  The breakdown of
these parametric errors is detailed in Table~\ref{tab:SMpred}, except for the contributions coming from the uncertainty in $m_H$, which, even in the conservative scenario, are numerically irrelevant in the total parametric uncertainty.

\begin{figure}[t]
  \centering
  \includegraphics[width=.5\textwidth]{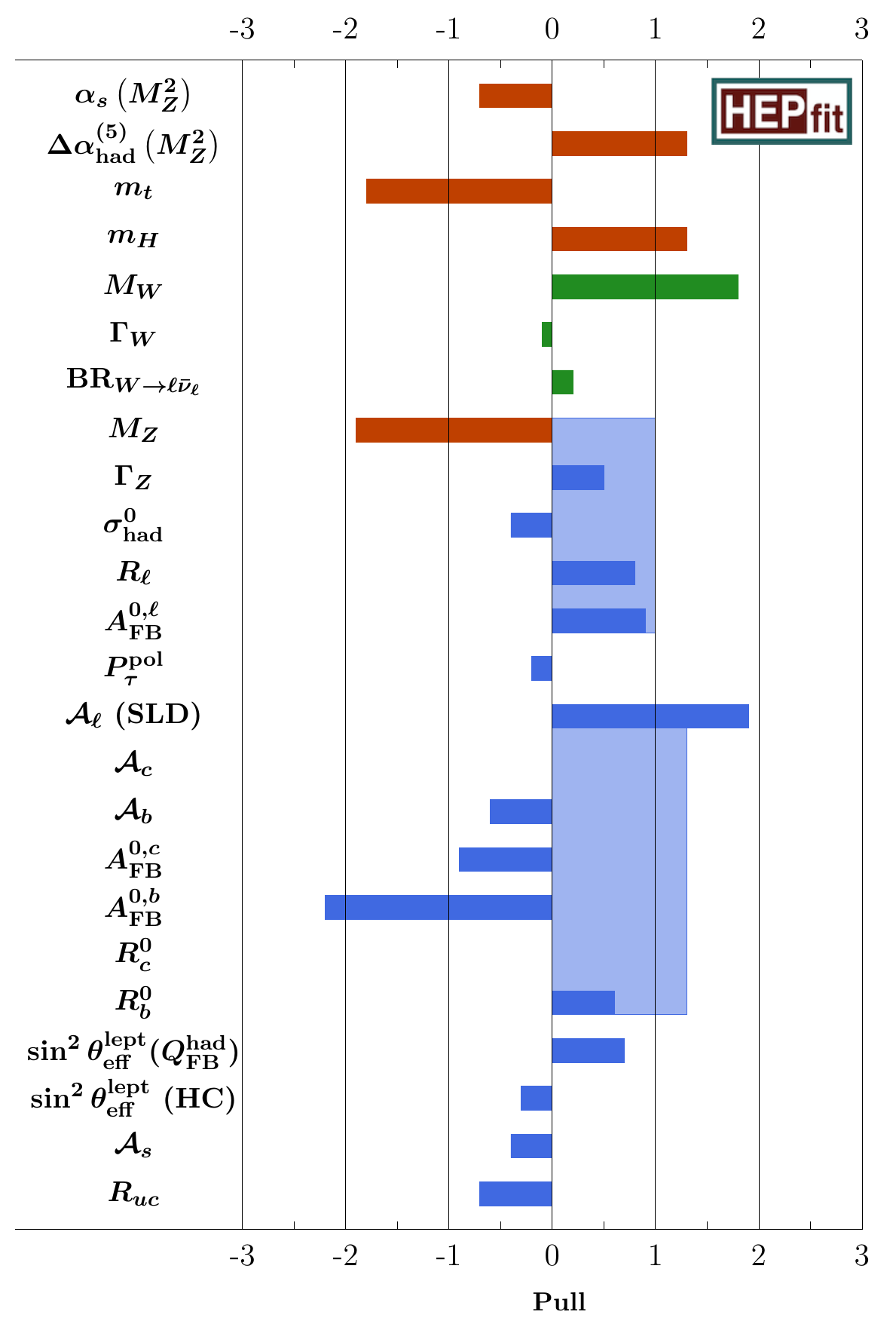}
  \caption{1D pulls between the observed experimental values and the SM predictions (indirect determinations) for the different EWPO (SM input parameters) considered in the fit, for the {\it standard} scenario. (The different colors in the figure are simply used to distinguish the SM inputs [orange], charged-current observables [green] and neutral-current observables [blue].) Each prediction is obtained removing the corresponding observable from the fit. The {\it transparent} bars represent the corresponding nD pulls for groups of correlated observables. See text for details. 
  }
  \label{fig:SMpulls}
\end{figure}

For each observable, we give in Tables~\ref{tab:SMfit-standard} and
\ref{tab:SMfit-conservative}, the experimental information used as
input (\textit{Measurement}), together with the output of the combined
fit (\textit{Posterior}), and the \textit{Prediction} of the same
quantity. The latter is obtained from the posterior predictive
distribution derived from a combined analysis of all the other
quantities. The compatibility of the constraints is then evaluated by sampling the posterior predictive distribution and the experimental one, by constructing the probability density function (p.d.f.) of the residuals $p(x)$, and by computing the integral of the p.d.f. in the region $p(x) < p(0)$. This two-sided $p-value$ is then converted to the equivalent number of standard
deviations for a Gaussian distribution. In the case of a Gaussian posterior predictive distribution, this quantity coincides with the
usual pull defined as the difference between the central values of
the two distributions divided by the sum in quadrature of the
residual mean square of the distributions themselves.
The advantage of this approach is that no approximation is made
on the shape of p.d.f.'s. 
These {\it 1D pulls} are also shown in Figure~\ref{fig:SMpulls}. 
We can see a clear consistency between
the measurement of all EWPO and their SM predictions. Only $A_{\rm FB}^{0, b}$ shows
some tension (at the $2\sigma$ level), which should be considered in
investigating new physics but also treated with care given the large
number of observables considered in the EW fit (see the discussion below for a quantitative assessment of the global significance taking the \emph{look-elsewhere} effect into account).

\begin{table}[t]
{
\begin{center}
\begin{tabular}{l|c|cc|cc}
\toprule
& Measurement & Full Indirect & Pull & Full Prediction & Pull \\
\hline 
\hline 
$\alpha_{s}(M_{Z}^2)$ & $ 0.1177 \pm 0.0010 $ & $ 0.1217 \pm 0.0046 $ & -0.8 & $ 0.1177 \pm 0.0010 $& -- \\ 
$\Delta\alpha^{(5)}_{\mathrm{had}}(M_{Z}^2)$ & $ 0.02766 \pm 0.00010 $ & $ 0.02752 \pm 0.00066 $ & 0.2 & $ 0.02766 \pm 0.00010 $& -- \\ 
$M_Z$ [GeV] & $ 91.1875 \pm 0.0021 $ & $ 91.200 \pm 0.039 $ & -0.3 & $ 91.1875 \pm 0.0021 $& -- \\ 
$m_t$ [GeV] & $ 172.58 \pm 0.45 $ & $ 180.1 \pm 9.6 $ & -0.8 & $ 172.58 \pm 0.45 $& -- \\ 
$m_H$ [GeV] & $ 125.21 \pm 0.12 $ & $ 196.2 \pm 89.9 $ & -0.4 & $ 125.21 \pm 0.12 $& -- \\ 
\hline 
$M_W$ [GeV] & $ 80.379 \pm 0.012 $ & $ 80.379 \pm 0.012 $ & 0.0 & $ 80.3545 \pm 0.0059 $& 1.8 \\ 
\hline 
$\Gamma_{W}$ [GeV] & $ 2.085 \pm 0.042 $ & $ 2.0916 \pm 0.0023 $ & -0.1 & $ 2.08782 \pm 0.00060 $& -0.1 \\ 
\hline 
$\sin^2\theta_{\rm eff}^{\rm lept}(Q_{\rm FB}^{\rm had})$ & $ 0.2324 \pm 0.0012 $ & $ 0.23147 \pm 0.00014 $ & 0.8 & $ 0.231534 \pm 0.000062 $& 0.7 \\ 
\hline 
$P_{\tau}^{\rm pol}=\mathcal{A}_\ell$ & $ 0.1465 \pm 0.0033 $ & $ 0.1474 \pm 0.0011 $ & -0.3 & $ 0.14692 \pm 0.00049 $& -0.1 \\ 
\hline 
$\Gamma_{Z}$ [GeV] & $ 2.4955 \pm 0.0023 $ & $ 2.4947 \pm 0.0020 $ & 0.3 & $ 2.49414 \pm 0.00068 $& 0.6 \\ 
$\sigma_{h}^{0}$ [nb] & $ 41.480 \pm 0.033 $ & $ 41.466 \pm 0.031 $ & 0.3 & $ 41.4929 \pm 0.0081 $& -0.4 \\ 
$R^{0}_{\ell}$ & $ 20.767 \pm 0.025 $ & $ 20.765 \pm 0.022 $ & 0.1 & $ 20.7464 \pm 0.0086 $& 0.8 \\ 
$A_{\rm FB}^{0, \ell}$ & $ 0.0171 \pm 0.0010 $ & $ 0.01630 \pm 0.00024 $ & 0.8 & $ 0.01619 \pm 0.00011 $& 0.9 \\ 
\hline 
$\mathcal{A}_{\ell}$ (SLD) & $ 0.1513 \pm 0.0021 $ & $ 0.1474 \pm 0.0011 $ & 1.6 & $ 0.14692 \pm 0.00049 $& 2.0 \\ 
$R^{0}_{b}$ & $ 0.21629 \pm 0.00066 $ & $ 0.21562 \pm 0.00035 $ & 0.9 & $ 0.21588 \pm 0.00010 $& 0.6 \\ 
$R^{0}_{c}$ & $ 0.1721 \pm 0.0030 $ & $ 0.17233 \pm 0.00017 $ & -0.1 & $ 0.172198 \pm 0.000054 $& 0.0 \\ 
$A_{\rm FB}^{0, b}$ & $ 0.0996 \pm 0.0016 $ & $ 0.10334 \pm 0.00077 $ & -2.1 & $ 0.10300 \pm 0.00034 $& -2.1 \\ 
$A_{\rm FB}^{0, c}$ & $ 0.0707 \pm 0.0035 $ & $ 0.07386 \pm 0.00059 $ & -0.9 & $ 0.07358 \pm 0.00026 $& -0.8 \\ 
$\mathcal{A}_b$ & $ 0.923 \pm 0.020 $ & $ 0.93468 \pm 0.00016 $ & -0.6 & $ 0.934727 \pm 0.000041 $& -0.6 \\ 
$\mathcal{A}_c$ & $ 0.670 \pm 0.027 $ & $ 0.66805 \pm 0.00048 $ & 0.1 & $ 0.66775 \pm 0.00023 $& 0.1 \\ 
\hline 
$\mathcal{A}_s$ & $ 0.895 \pm 0.091 $ & $ 0.935693 \pm 0.000088 $ & -0.4 & $ 0.935637 \pm 0.000041 $& -0.4 \\ 
BR$_{W\to\ell\bar\nu_\ell}$ & $ 0.10860 \pm 0.00090 $ & $ 0.10829 \pm 0.00011 $ & 0.3 & $ 0.108386 \pm 0.000023 $& 0.2 \\ 
$\sin^2\theta_{\rm eff}^{\rm lept}$ (HC) & $ 0.23143 \pm 0.00025 $ & $ 0.23147 \pm 0.00014 $ & -0.1 & $ 0.231534 \pm 0.000062 $& -0.4 \\ 
\hline 
$R_{uc}$ & $ 0.1660 \pm 0.0090 $ & $ 0.17236 \pm 0.00017 $ & -0.7 & $ 0.172220 \pm 0.000032 $& -0.7 \\ 
\bottomrule
\end{tabular}
\end{center}
}
\caption{Results of the \emph{full indirect} determination of SM parameters using only EWPD (third column) and of the \emph{full prediction} for EWPO using only information on SM parameters (fourth column). For comparison, the input values are reported in the second column. See the text for details.}
\label{tab:FullIndFullPred}
\end{table}
\begin{figure}[hb!]
  \centering
  \includegraphics[width=.45\textwidth]{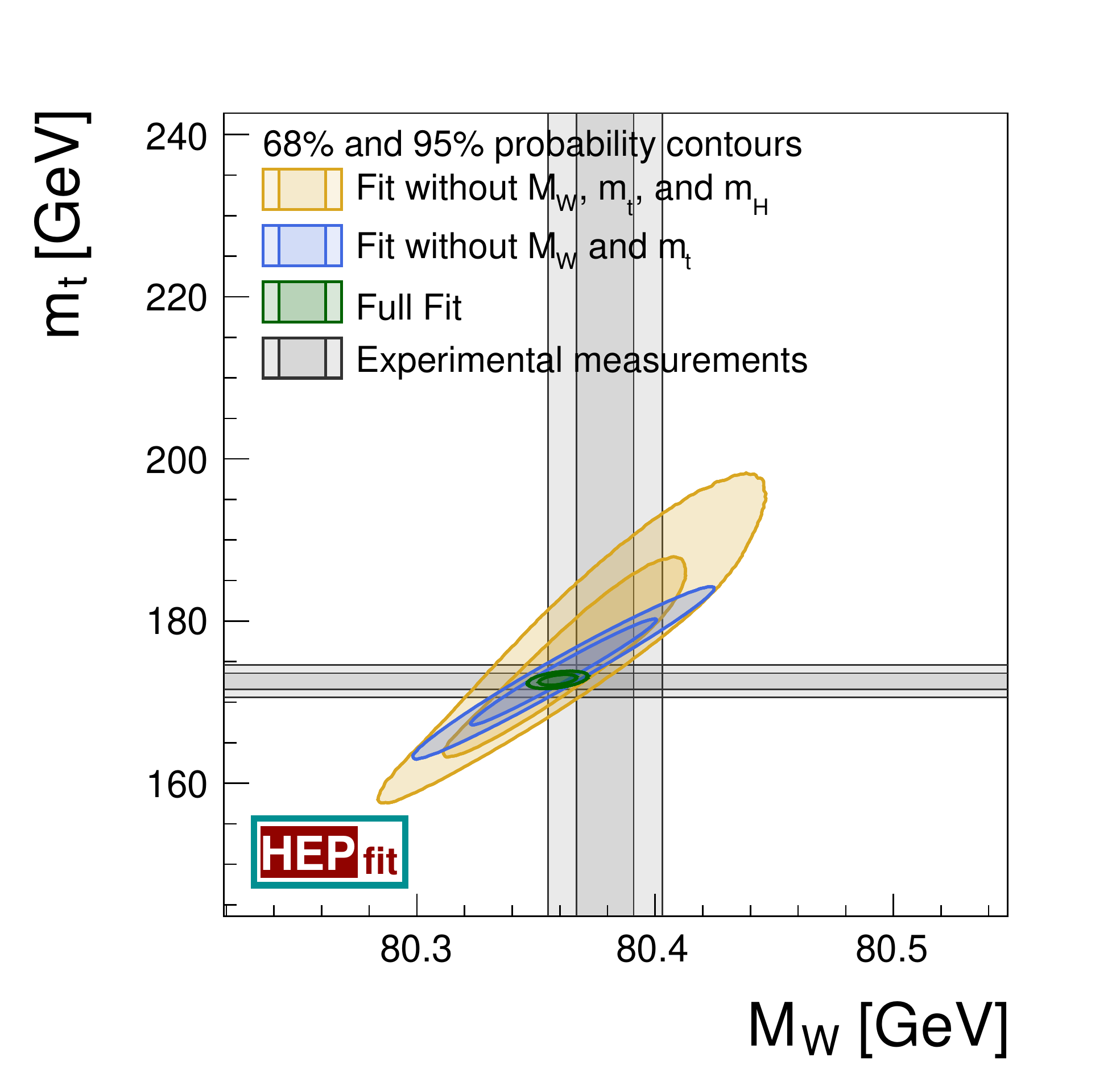}
  \hspace{-2mm}
  \includegraphics[width=.45\textwidth]{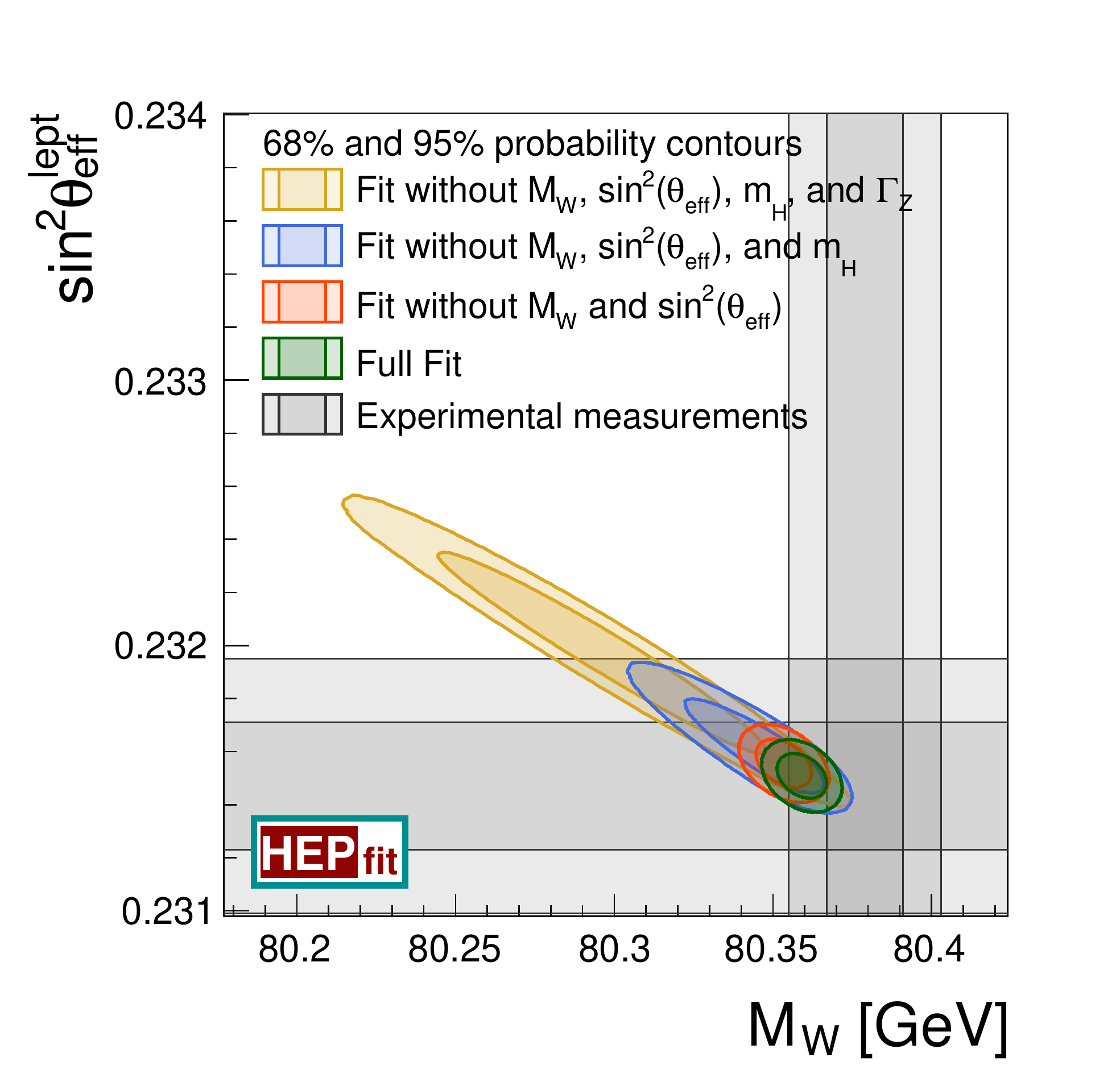}
  \caption{Impact of various constraints in the  $m_t$ vs. $M_W$ (left) and $\sin^2{\theta_{\rm eff}^{\rm lept}}$ vs. $M_W$ (right) planes. Dark (light)
    regions correspond to $68\%$ ($95\%$) probability ranges.}
  \label{fig:2Dplots}
\end{figure}

Moreover, when interpreting this {\it 1D pull} one needs to take into account 
that $A_{\rm FB}^{0, b}$ is actually part of a set of experimentally correlated observables.
In order to check the consistency between SM and experiments in this case, one can
define an \emph{nD pull}
by removing from the fit one set of correlated observables at a time and
computing the prediction for the set of observables together with their
covariance matrix. Then the same procedure described for \emph{1D pulls} can be carried out, this time sampling the posterior predictive and experimental n-dimensional p.d.f.'s. This {\emph{nD pull}} is shown in
the last column in Tables~\ref{tab:SMfit-standard} and \ref{tab:SMfit-conservative}, as well as in Figure~\ref{fig:SMpulls}, 
in which case we
see that the global pull for the set of correlated observables involving $A_{\rm FB}^{0, b}$ is reduced to $1.3~\!\sigma$. To get an idea of the agreement between the SM and EWPD, it is useful to consider the distribution of the $p$-values corresponding to the 1D pulls for the individual measurements. For purely statistical fluctuations, one expects the $p$-values to be uniformly distributed between 0 and 1. From the results in Tables \ref{tab:SMfit-standard} and \ref{tab:SMfit-conservative}, we obtain in both scenarios an average $p$-value of $0.5$ with $\sigma = 0.3$, fully compatible with a flat distribution. 

\begin{figure}[t!]
  \centering
  \includegraphics[width=.345\textwidth]{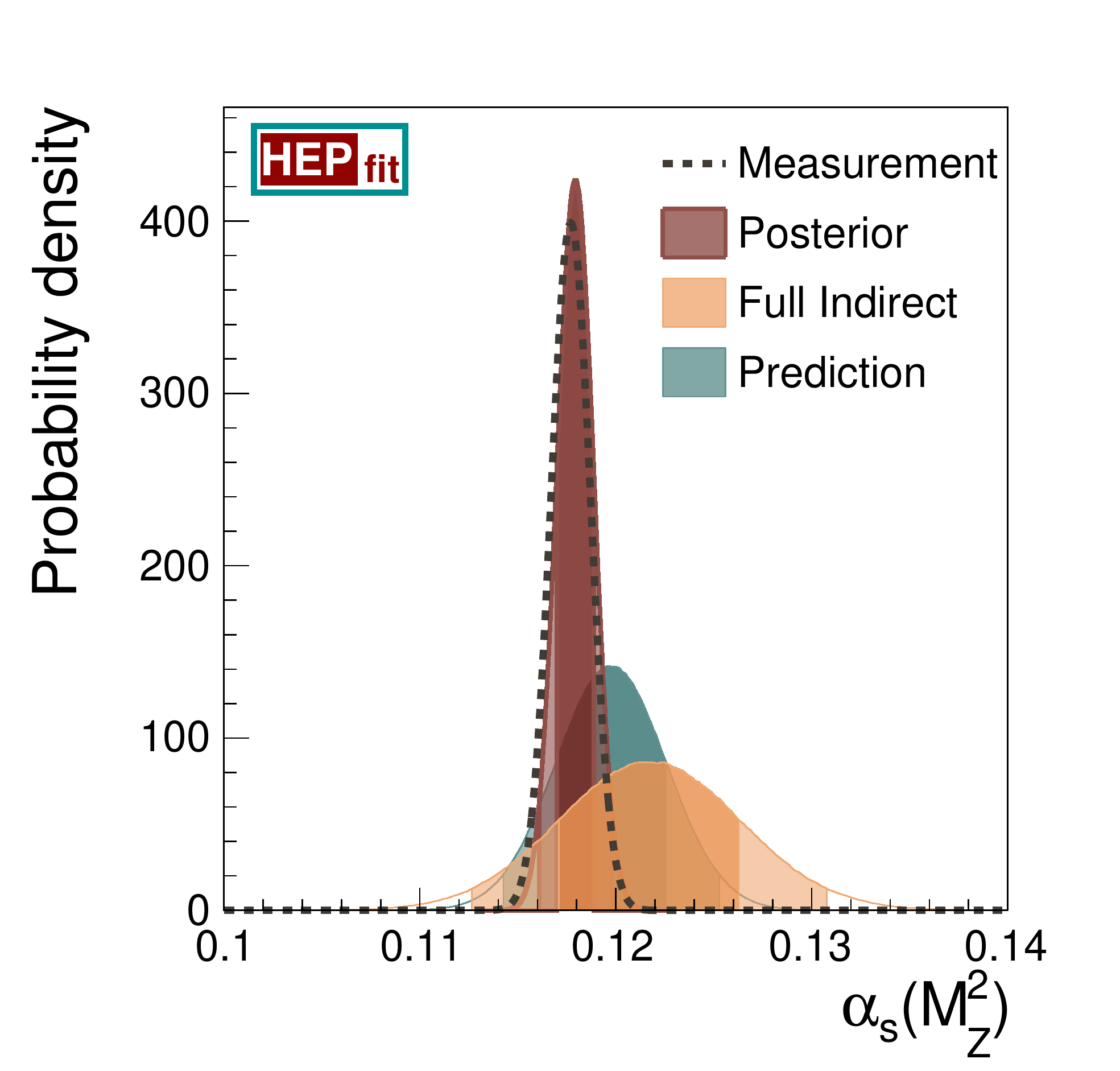}
  \hspace{-6.3mm}
  \includegraphics[width=.345\textwidth]{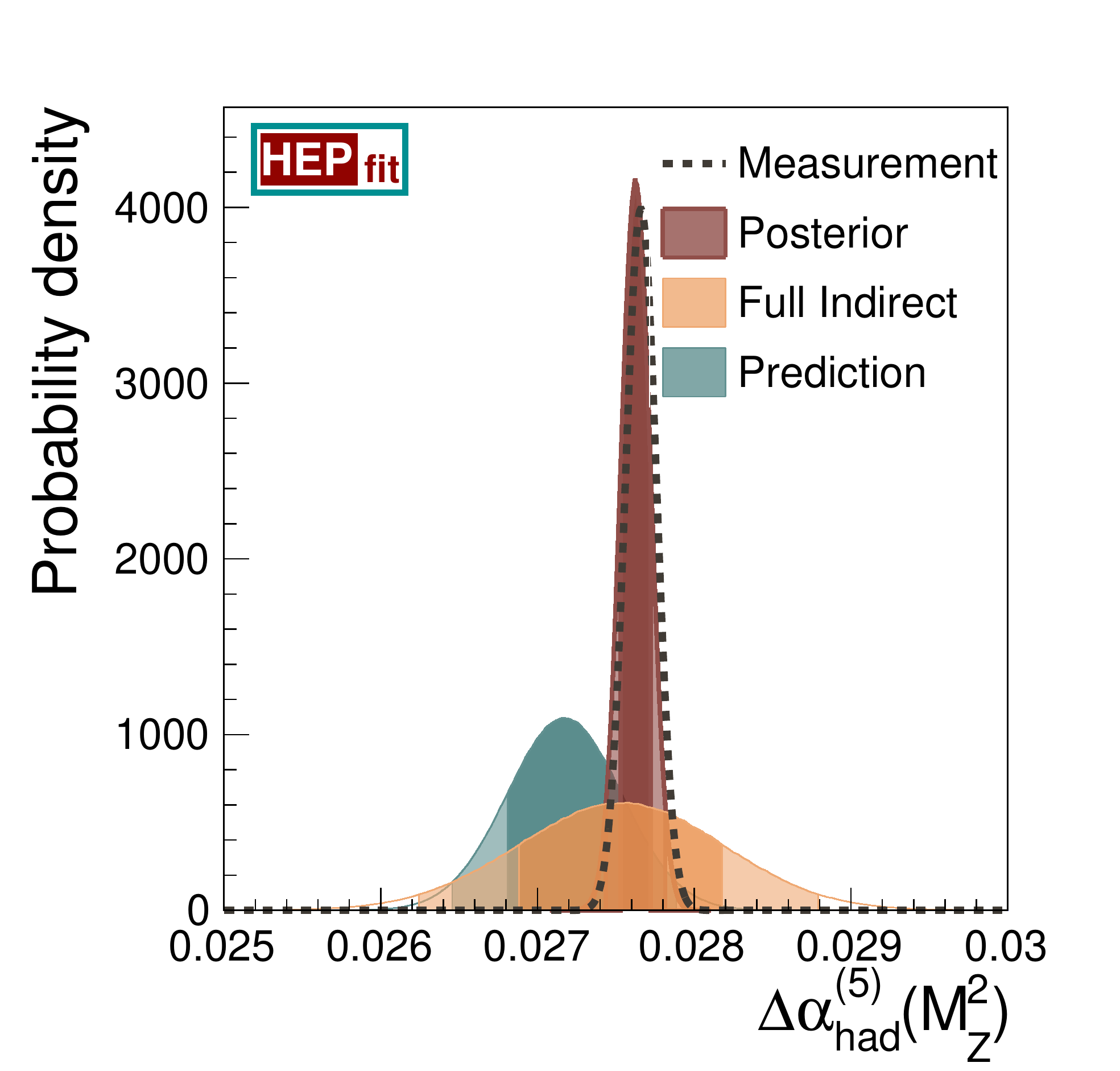}
  \hspace{-6.3mm}
  \includegraphics[width=.345\textwidth]{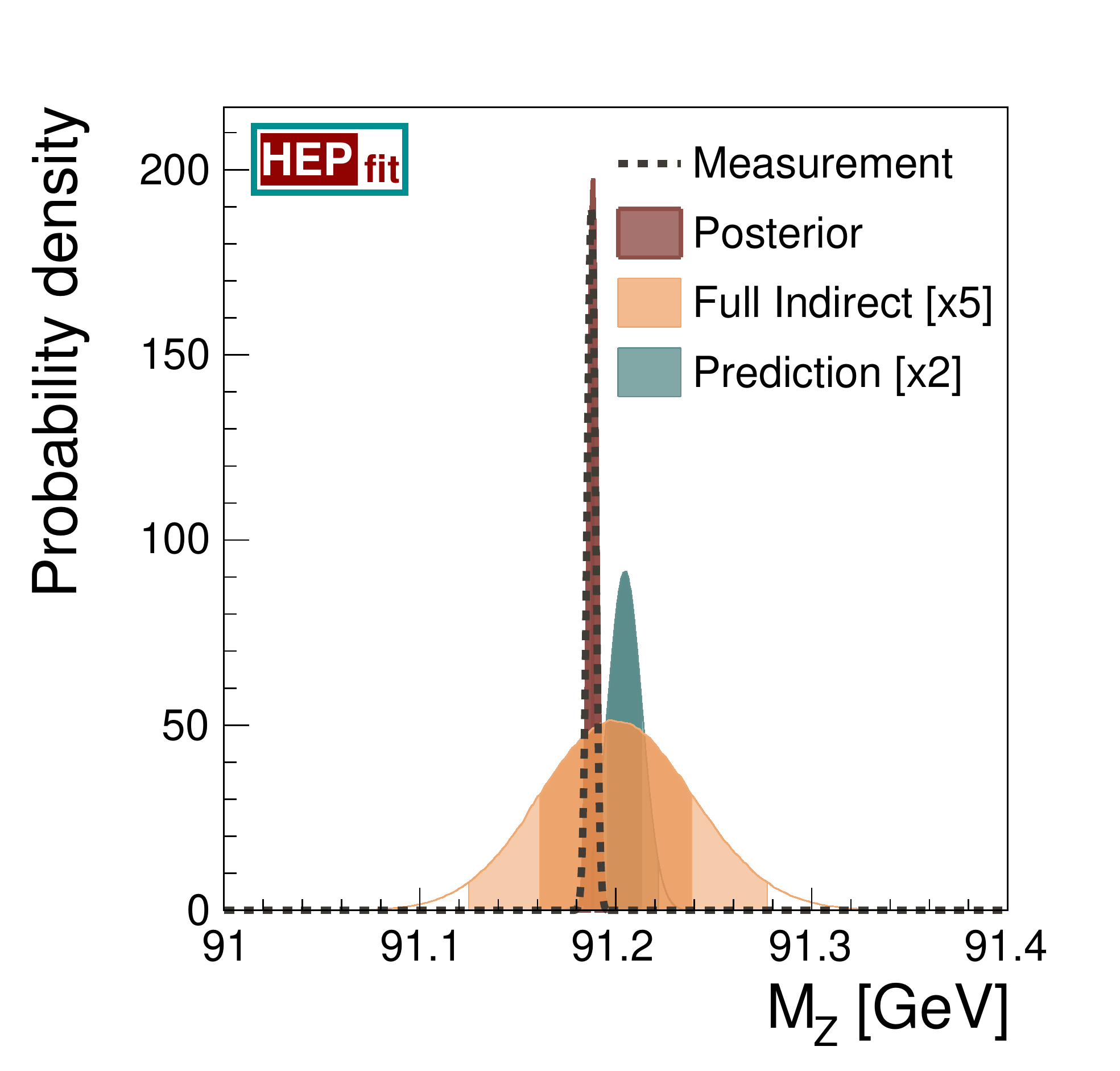}
  \\
  \includegraphics[width=.345\textwidth]{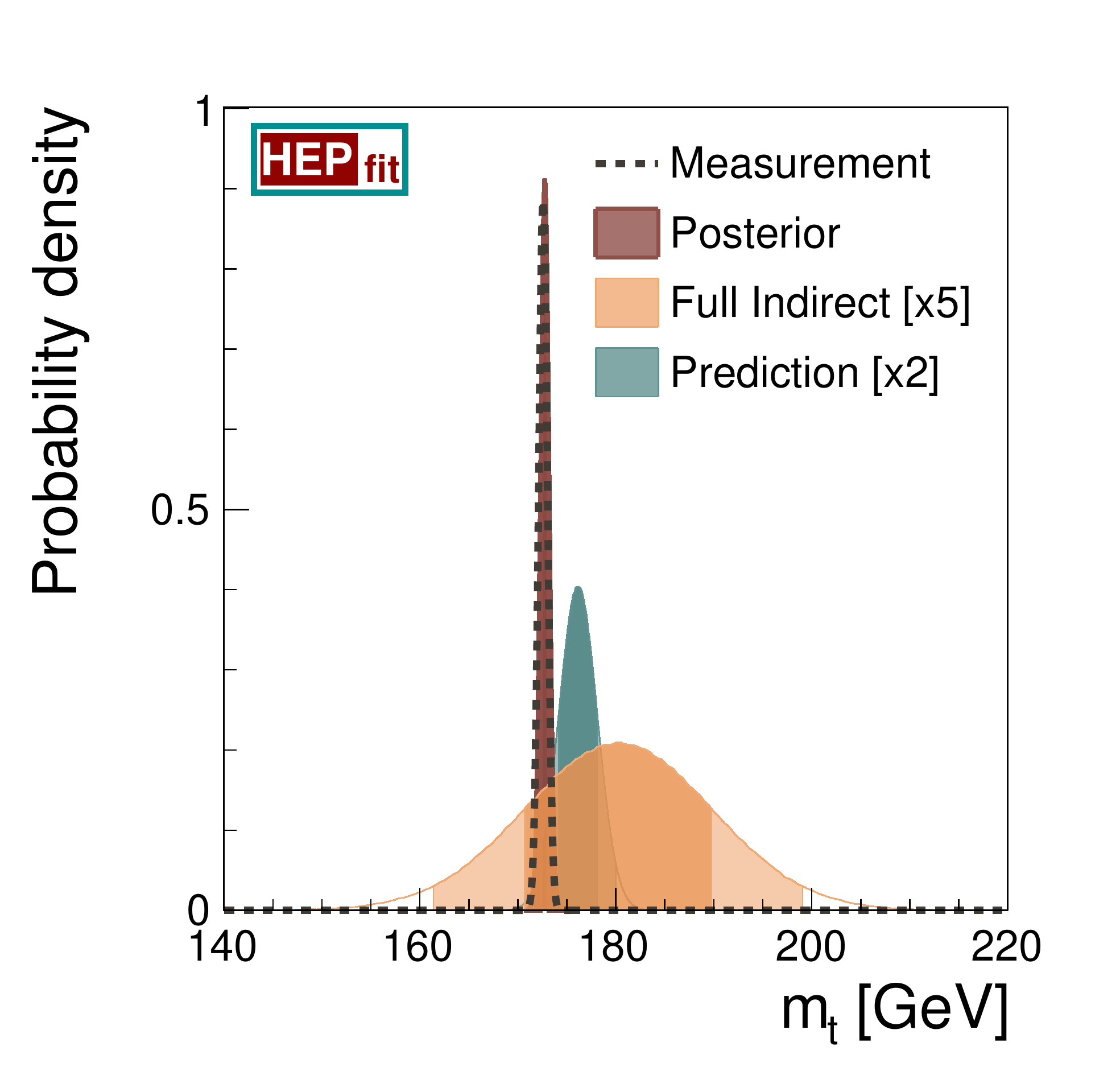}
  \hspace{-6.3mm}
  \includegraphics[width=.345\textwidth]{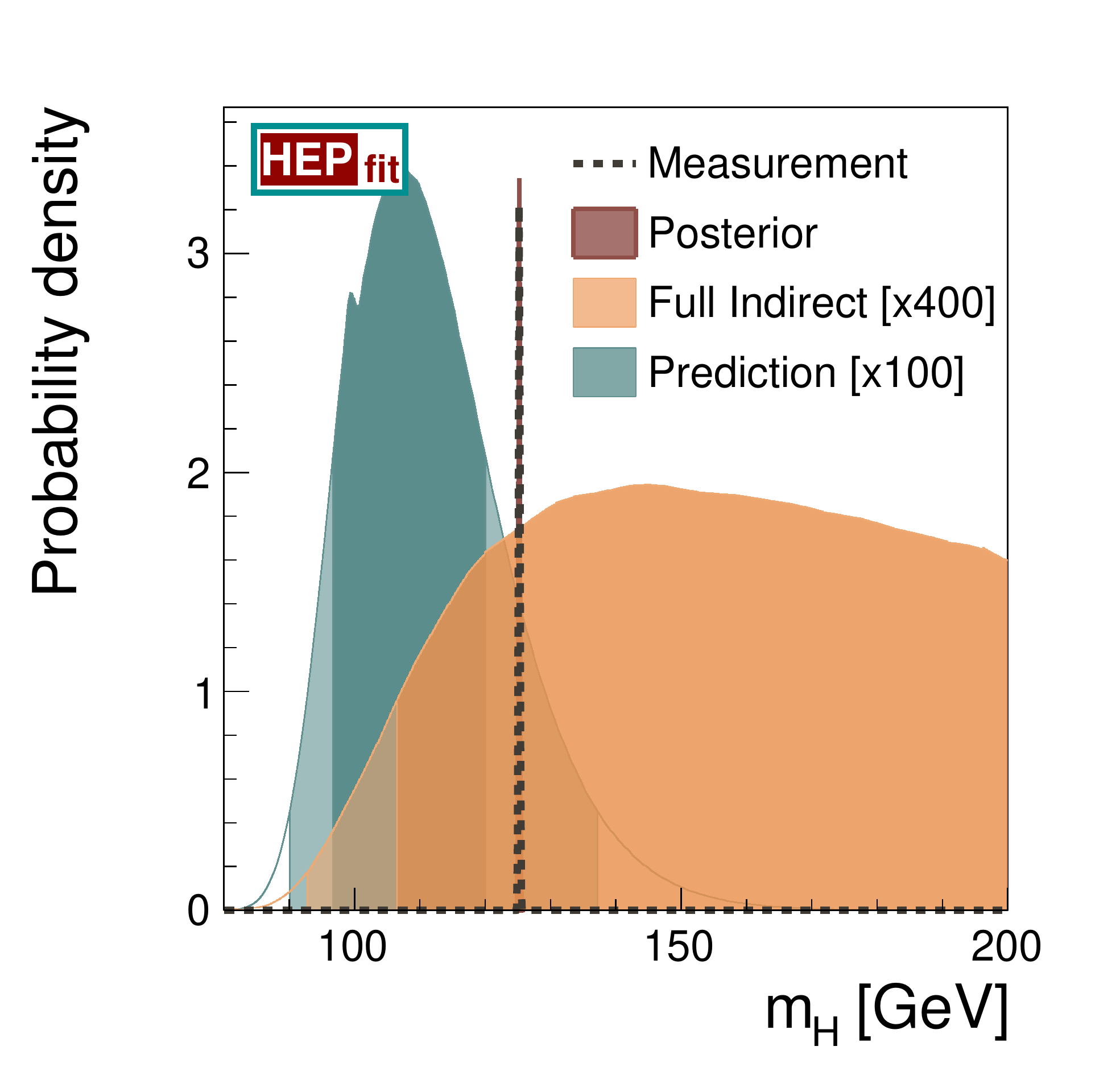}
  \caption{Comparison among the direct measurement, the posterior,
    the posterior predictive (or indirect) probability distribution (denoted by "Prediction"), and the full indirect determination
    of the input parameters in the SM fit. The posterior predictive and the full indirect determination distributions are obtained
    from the fit by assuming a flat prior for the parameter under
    consideration or for all SM parameters respectively. To allow for a comparison with the posterior predictive distribution, the full indirect p.d.f for the Higgs mass is truncated in the figure. Dark (light)
    regions correspond to $68\%$ ($95\%$) probability ranges. {\tt HEPfit}  uses different semi-analytical approximations to the full available calculations of the EWPD, depending the range of variation of the SM inputs, see e.g.~\cite{Dubovyk:2018rlg}. The small ``bump" in the posterior predictive of the $m_H$ figure arises as a result of the transition between these approximated expressions.
    }
  \label{fig:SMinputs}
\end{figure}

In addition to the \emph{Predictions} obtained removing each individual observable from the fit, one can obtain a \emph{Full Prediction} by dropping all experimental information on EWPO and just using the SM and the information on SM parameters. Conversely, one can obtain a \emph{Full Indirect} determination of the SM parameters by dropping all information on all parameters simultaneously and determining all of them from the fit to EWPD. The results of these two extreme possibilities are reported in Table \ref{tab:FullIndFullPred}. Again, for the \emph{Full Prediction} case we obtain an average $p$-value of 0.5 with $\sigma = 0.3$, fully compatible with a flat distribution. The results of the \emph{Full Indirect} fit represent the best possible agreement with EWPD one could possibly reach in the SM. Indeed, in this case the tension on $M_W$ disappears, while the tensions on $\mathcal{A}_\ell$ (SLD) and $A^{0,b}_\mathrm{FB}$ are only mildly decreased. One would need to go beyond the SM to improve the agreement further. The tension on $M_W$ is released by allowing for a larger value of $m_t$, as can be seen from the left panel of Fig.~\ref{fig:2Dplots}, where the impact of different constraints in shaping the two-dimensional p.d.f.'s of $m_t$ vs. $M_W$ and of $\sin^2{\theta_{\rm eff}^{\rm lept}}$ vs. $M_W$ is shown.

The p.d.f.'s for the SM input parameters are reported in Fig.~\ref{fig:SMinputs}, together with the posterior from the fit, the indirect determination and the \emph{full indirect} one. While direct measurements are more precise than indirect determinations (by orders of magnitude in the case of the Higgs mass), all indirect determinations are compatible with the measurements within $2 \sigma$. Our indirect determination of $\Delta \alpha_\mathrm{had}^{(5)}(M_Z^2)$ fully agrees with the independent one recently obtained in Ref.~\cite{Keshavarzi:2020bfy} using the Gfitter library \cite{Flacher:2008zq}.
The pull between the determination of $\Delta \alpha_\mathrm{had}^{(5)}(M_Z^2)$ based on the BMW lattice calculation~\cite{Borsanyi:2017zdw} and the indirect determination is of $1.3\sigma$ ($1.1\sigma$) in the \textit{standard} (\textit{conservative}) scenario, showing no inconsistency between the current lattice evaluation and the EW fit. It will be very interesting to see if the good agreement between the lattice determination and the EW fit persists when the updated lattice value corresponding to the value of the hadronic vacuum polarization recently published in Ref.~\cite{Borsanyi:2020mff} is released.
The indirect determination of the top mass is also compatible with the measurement at less than $2 \sigma$, but on the larger $m_t$ side, bringing the SM further away from the Planck stability bound (see e.g.~Ref.~\cite{Buttazzo:2013uya}).

In conclusion, EWPD appear to be fully compatible with the SM, with no more tensions than expected from statistical fluctuations. In the \emph{standard scenario} SM fit, the largest pull neglecting correlations is $2.2\sigma$ on 24 observables, while taking correlations into account it is $1.8\sigma$ on 14 observables. In both the \emph{full indirect} and \emph{full prediction} determinations, the largest pull neglecting correlations is $2.1\sigma$ on 24 observables. To quantify further the agreement of the SM, we generated 600 toy experiments centered on the \emph{full prediction} with the current experimental uncertainty and computed the fraction of toys in which the largest pull was larger than the largest one observed in real data. This fraction is an estimate of the \emph{global p-value}. Neglecting correlations, we obtain $p=0.53$, corresponding to $0.6\sigma$ for a Gaussian distribution, while taking into account the correlations (fixed to the values observed in current data) we get $p=0.45$, corresponding to $0.8\sigma$.

\section*{Acknowledgements}
We thank Laurent Lellouch for informative discussions on the results of Ref.~\cite{Borsanyi:2017zdw}.
This work was supported in part by the Italian Ministry of Research (MIUR) under grant PRIN 20172LNEEZ.
The work of J.B. has been supported by the FEDER/Junta de Andaluc\'ia project grant P18-FRJ-3735. The work of S.M. has been supported by the Japan Society for the Promotion of Science under grant 17K05429. The work of L.R and A.G. has been supported by the U.S. Department of Energy under grant DE-SC0010102.

\bibliography{hepbiblio}

\begin{thebibliography}{76}%
\makeatletter
\providecommand \@ifxundefined [1]{%
 \@ifx{#1\undefined}
}%
\providecommand \@ifnum [1]{%
 \ifnum #1\expandafter \@firstoftwo
 \else \expandafter \@secondoftwo
 \fi
}%
\providecommand \@ifx [1]{%
 \ifx #1\expandafter \@firstoftwo
 \else \expandafter \@secondoftwo
 \fi
}%
\providecommand \natexlab [1]{#1}%
\providecommand \enquote  [1]{``#1''}%
\providecommand \bibnamefont  [1]{#1}%
\providecommand \bibfnamefont [1]{#1}%
\providecommand \citenamefont [1]{#1}%
\providecommand \href@noop [0]{\@secondoftwo}%
\providecommand \href [0]{\begingroup \@sanitize@url \@href}%
\providecommand \@href[1]{\@@startlink{#1}\@@href}%
\providecommand \@@href[1]{\endgroup#1\@@endlink}%
\providecommand \@sanitize@url [0]{\catcode `\\12\catcode `\$12\catcode
  `\&12\catcode `\#12\catcode `\^12\catcode `\_12\catcode `\%12\relax}%
\providecommand \@@startlink[1]{}%
\providecommand \@@endlink[0]{}%
\providecommand \url  [0]{\begingroup\@sanitize@url \@url }%
\providecommand \@url [1]{\endgroup\@href {#1}{\urlprefix }}%
\providecommand \urlprefix  [0]{URL }%
\providecommand \Eprint [0]{\href }%
\providecommand \doibase [0]{https://doi.org/}%
\providecommand \selectlanguage [0]{\@gobble}%
\providecommand \bibinfo  [0]{\@secondoftwo}%
\providecommand \bibfield  [0]{\@secondoftwo}%
\providecommand \translation [1]{[#1]}%
\providecommand \BibitemOpen [0]{}%
\providecommand \bibitemStop [0]{}%
\providecommand \bibitemNoStop [0]{.\EOS\space}%
\providecommand \EOS [0]{\spacefactor3000\relax}%
\providecommand \BibitemShut  [1]{\csname bibitem#1\endcsname}%
\let\auto@bib@innerbib\@empty
\bibitem [{\citenamefont {Sirlin}(1980)}]{Sirlin:1980nh}%
  \BibitemOpen
  \bibfield  {author} {\bibinfo {author} {\bibfnamefont {A.}~\bibnamefont
  {Sirlin}},\ }\bibfield  {title} {\bibinfo {title} {{Radiative corrections in
  the $SU(2)_L\times U(1)$ theory: A simple renormalization framework}},\
  }\href {https://doi.org/10.1103/PhysRevD.22.971} {\bibfield  {journal}
  {\bibinfo  {journal} {Phys.Rev.}\ }\textbf {\bibinfo {volume} {D22}},\
  \bibinfo {pages} {971} (\bibinfo {year} {1980})}\BibitemShut {NoStop}%
\bibitem [{\citenamefont {Marciano}\ and\ \citenamefont
  {Sirlin}(1980)}]{Marciano:1980pb}%
  \BibitemOpen
  \bibfield  {author} {\bibinfo {author} {\bibfnamefont {W.}~\bibnamefont
  {Marciano}}\ and\ \bibinfo {author} {\bibfnamefont {A.}~\bibnamefont
  {Sirlin}},\ }\bibfield  {title} {\bibinfo {title} {{Radiative corrections to
  neutrino induced neutral current phenomena in the $SU(2)_L\times U(1)$
  theory}},\ }\href {https://doi.org/10.1103/PhysRevD.31.213,
  10.1103/PhysRevD.22.2695} {\bibfield  {journal} {\bibinfo  {journal}
  {Phys.Rev.}\ }\textbf {\bibinfo {volume} {D22}},\ \bibinfo {pages} {2695}
  (\bibinfo {year} {1980})}\BibitemShut {NoStop}%
\bibitem [{\citenamefont {Djouadi}\ and\ \citenamefont
  {Verzegnassi}(1987)}]{Djouadi:1987gn}%
  \BibitemOpen
  \bibfield  {author} {\bibinfo {author} {\bibfnamefont {A.}~\bibnamefont
  {Djouadi}}\ and\ \bibinfo {author} {\bibfnamefont {C.}~\bibnamefont
  {Verzegnassi}},\ }\bibfield  {title} {\bibinfo {title} {{Virtual very heavy
  top effects in LEP/SLC precision measurements}},\ }\href
  {https://doi.org/10.1016/0370-2693(87)91206-8} {\bibfield  {journal}
  {\bibinfo  {journal} {Phys.Lett.}\ }\textbf {\bibinfo {volume} {B195}},\
  \bibinfo {pages} {265} (\bibinfo {year} {1987})}\BibitemShut {NoStop}%
\bibitem [{\citenamefont {Djouadi}(1988)}]{Djouadi:1987di}%
  \BibitemOpen
  \bibfield  {author} {\bibinfo {author} {\bibfnamefont {A.}~\bibnamefont
  {Djouadi}},\ }\bibfield  {title} {\bibinfo {title} {{${\cal
  O}(\alpha\alpha_s)$ vacuum polarization functions of the standard model gauge
  bosons}},\ }\href {https://doi.org/10.1007/BF02812964} {\bibfield  {journal}
  {\bibinfo  {journal} {Nuovo Cim.}\ }\textbf {\bibinfo {volume} {A100}},\
  \bibinfo {pages} {357} (\bibinfo {year} {1988})}\BibitemShut {NoStop}%
\bibitem [{\citenamefont {Kniehl}(1990)}]{Kniehl:1989yc}%
  \BibitemOpen
  \bibfield  {author} {\bibinfo {author} {\bibfnamefont {B.~A.}\ \bibnamefont
  {Kniehl}},\ }\bibfield  {title} {\bibinfo {title} {{Two-loop corrections to
  the vacuum polarizations in perturbative QCD}},\ }\href
  {https://doi.org/10.1016/0550-3213(90)90552-O} {\bibfield  {journal}
  {\bibinfo  {journal} {Nucl.Phys.}\ }\textbf {\bibinfo {volume} {B347}},\
  \bibinfo {pages} {86} (\bibinfo {year} {1990})}\BibitemShut {NoStop}%
\bibitem [{\citenamefont {Halzen}\ and\ \citenamefont
  {Kniehl}(1991)}]{Halzen:1990je}%
  \BibitemOpen
  \bibfield  {author} {\bibinfo {author} {\bibfnamefont {F.}~\bibnamefont
  {Halzen}}\ and\ \bibinfo {author} {\bibfnamefont {B.~A.}\ \bibnamefont
  {Kniehl}},\ }\bibfield  {title} {\bibinfo {title} {{$\Delta r$ beyond one
  loop}},\ }\href {https://doi.org/10.1016/0550-3213(91)90319-S} {\bibfield
  {journal} {\bibinfo  {journal} {Nucl.Phys.}\ }\textbf {\bibinfo {volume}
  {B353}},\ \bibinfo {pages} {567} (\bibinfo {year} {1991})}\BibitemShut
  {NoStop}%
\bibitem [{\citenamefont {Kniehl}\ and\ \citenamefont
  {Sirlin}(1992)}]{Kniehl:1991gu}%
  \BibitemOpen
  \bibfield  {author} {\bibinfo {author} {\bibfnamefont {B.~A.}\ \bibnamefont
  {Kniehl}}\ and\ \bibinfo {author} {\bibfnamefont {A.}~\bibnamefont
  {Sirlin}},\ }\bibfield  {title} {\bibinfo {title} {{Dispersion relations for
  vacuum polarization functions in electroweak physics}},\ }\href
  {https://doi.org/10.1016/0550-3213(92)90232-Z} {\bibfield  {journal}
  {\bibinfo  {journal} {Nucl.Phys.}\ }\textbf {\bibinfo {volume} {B371}},\
  \bibinfo {pages} {141} (\bibinfo {year} {1992})}\BibitemShut {NoStop}%
\bibitem [{\citenamefont {Kniehl}\ and\ \citenamefont
  {Sirlin}(1993)}]{Kniehl:1992dx}%
  \BibitemOpen
  \bibfield  {author} {\bibinfo {author} {\bibfnamefont {B.~A.}\ \bibnamefont
  {Kniehl}}\ and\ \bibinfo {author} {\bibfnamefont {A.}~\bibnamefont
  {Sirlin}},\ }\bibfield  {title} {\bibinfo {title} {{On the effect of the $t
  \bar{t}$ threshold on electroweak parameters}},\ }\href
  {https://doi.org/10.1103/PhysRevD.47.883} {\bibfield  {journal} {\bibinfo
  {journal} {Phys.Rev.}\ }\textbf {\bibinfo {volume} {D47}},\ \bibinfo {pages}
  {883} (\bibinfo {year} {1993})}\BibitemShut {NoStop}%
\bibitem [{\citenamefont {Barbieri}\ \emph {et~al.}(1992)\citenamefont
  {Barbieri}, \citenamefont {Beccaria}, \citenamefont {Ciafaloni},
  \citenamefont {Curci},\ and\ \citenamefont {Vicere}}]{Barbieri:1992nz}%
  \BibitemOpen
  \bibfield  {author} {\bibinfo {author} {\bibfnamefont {R.}~\bibnamefont
  {Barbieri}}, \bibinfo {author} {\bibfnamefont {M.}~\bibnamefont {Beccaria}},
  \bibinfo {author} {\bibfnamefont {P.}~\bibnamefont {Ciafaloni}}, \bibinfo
  {author} {\bibfnamefont {G.}~\bibnamefont {Curci}},\ and\ \bibinfo {author}
  {\bibfnamefont {A.}~\bibnamefont {Vicere}},\ }\bibfield  {title} {\bibinfo
  {title} {{Radiative correction effects of a very heavy top}},\ }\href
  {https://doi.org/10.1016/0370-2693(92)91960-H} {\bibfield  {journal}
  {\bibinfo  {journal} {Phys.Lett.}\ }\textbf {\bibinfo {volume} {B288}},\
  \bibinfo {pages} {95} (\bibinfo {year} {1992})},\ \Eprint
  {https://arxiv.org/abs/hep-ph/9205238} {arXiv:hep-ph/9205238 [hep-ph]}
  \BibitemShut {NoStop}%
\bibitem [{\citenamefont {Barbieri}\ \emph {et~al.}(1993)\citenamefont
  {Barbieri}, \citenamefont {Beccaria}, \citenamefont {Ciafaloni},
  \citenamefont {Curci},\ and\ \citenamefont {Vicere}}]{Barbieri:1992dq}%
  \BibitemOpen
  \bibfield  {author} {\bibinfo {author} {\bibfnamefont {R.}~\bibnamefont
  {Barbieri}}, \bibinfo {author} {\bibfnamefont {M.}~\bibnamefont {Beccaria}},
  \bibinfo {author} {\bibfnamefont {P.}~\bibnamefont {Ciafaloni}}, \bibinfo
  {author} {\bibfnamefont {G.}~\bibnamefont {Curci}},\ and\ \bibinfo {author}
  {\bibfnamefont {A.}~\bibnamefont {Vicere}},\ }\bibfield  {title} {\bibinfo
  {title} {{Two-loop heavy top effects in the standard model}},\ }\href
  {https://doi.org/10.1016/0550-3213(93)90448-X} {\bibfield  {journal}
  {\bibinfo  {journal} {Nucl.Phys.}\ }\textbf {\bibinfo {volume} {B409}},\
  \bibinfo {pages} {105} (\bibinfo {year} {1993})}\BibitemShut {NoStop}%
\bibitem [{\citenamefont {Djouadi}\ and\ \citenamefont
  {Gambino}(1994)}]{Djouadi:1993ss}%
  \BibitemOpen
  \bibfield  {author} {\bibinfo {author} {\bibfnamefont {A.}~\bibnamefont
  {Djouadi}}\ and\ \bibinfo {author} {\bibfnamefont {P.}~\bibnamefont
  {Gambino}},\ }\bibfield  {title} {\bibinfo {title} {{Electroweak gauge bosons
  selfenergies: Complete QCD corrections}},\ }\href
  {https://doi.org/10.1103/PhysRevD.49.3499, 10.1103/PhysRevD.53.4111}
  {\bibfield  {journal} {\bibinfo  {journal} {Phys.Rev.}\ }\textbf {\bibinfo
  {volume} {D49}},\ \bibinfo {pages} {3499} (\bibinfo {year} {1994})},\ \Eprint
  {https://arxiv.org/abs/hep-ph/9309298} {arXiv:hep-ph/9309298 [hep-ph]}
  \BibitemShut {NoStop}%
\bibitem [{\citenamefont {Fleischer}\ \emph {et~al.}(1993)\citenamefont
  {Fleischer}, \citenamefont {Tarasov},\ and\ \citenamefont
  {Jegerlehner}}]{Fleischer:1993ub}%
  \BibitemOpen
  \bibfield  {author} {\bibinfo {author} {\bibfnamefont {J.}~\bibnamefont
  {Fleischer}}, \bibinfo {author} {\bibfnamefont {O.}~\bibnamefont {Tarasov}},\
  and\ \bibinfo {author} {\bibfnamefont {F.}~\bibnamefont {Jegerlehner}},\
  }\bibfield  {title} {\bibinfo {title} {{Two-loop heavy top corrections to the
  $\rho$ parameter: A simple formula valid for arbitrary Higgs mass}},\ }\href
  {https://doi.org/10.1016/0370-2693(93)90810-5} {\bibfield  {journal}
  {\bibinfo  {journal} {Phys.Lett.}\ }\textbf {\bibinfo {volume} {B319}},\
  \bibinfo {pages} {249} (\bibinfo {year} {1993})}\BibitemShut {NoStop}%
\bibitem [{\citenamefont {Fleischer}\ \emph {et~al.}(1995)\citenamefont
  {Fleischer}, \citenamefont {Tarasov},\ and\ \citenamefont
  {Jegerlehner}}]{Fleischer:1994cb}%
  \BibitemOpen
  \bibfield  {author} {\bibinfo {author} {\bibfnamefont {J.}~\bibnamefont
  {Fleischer}}, \bibinfo {author} {\bibfnamefont {O.}~\bibnamefont {Tarasov}},\
  and\ \bibinfo {author} {\bibfnamefont {F.}~\bibnamefont {Jegerlehner}},\
  }\bibfield  {title} {\bibinfo {title} {{Two-loop large top mass corrections
  to electroweak parameters: Analytic results valid for arbitrary Higgs
  mass}},\ }\href {https://doi.org/10.1103/PhysRevD.51.3820} {\bibfield
  {journal} {\bibinfo  {journal} {Phys.Rev.}\ }\textbf {\bibinfo {volume}
  {D51}},\ \bibinfo {pages} {3820} (\bibinfo {year} {1995})}\BibitemShut
  {NoStop}%
\bibitem [{\citenamefont {Avdeev}\ \emph {et~al.}(1994)\citenamefont {Avdeev},
  \citenamefont {Fleischer}, \citenamefont {Mikhailov},\ and\ \citenamefont
  {Tarasov}}]{Avdeev:1994db}%
  \BibitemOpen
  \bibfield  {author} {\bibinfo {author} {\bibfnamefont {L.}~\bibnamefont
  {Avdeev}}, \bibinfo {author} {\bibfnamefont {J.}~\bibnamefont {Fleischer}},
  \bibinfo {author} {\bibfnamefont {S.}~\bibnamefont {Mikhailov}},\ and\
  \bibinfo {author} {\bibfnamefont {O.}~\bibnamefont {Tarasov}},\ }\bibfield
  {title} {\bibinfo {title} {{${\cal O}(\alpha\alpha_s^2)$ correction to the
  electroweak $\rho$ parameter}},\ }\href
  {https://doi.org/10.1016/0370-2693(94)90573-8} {\bibfield  {journal}
  {\bibinfo  {journal} {Phys.Lett.}\ }\textbf {\bibinfo {volume} {B336}},\
  \bibinfo {pages} {560} (\bibinfo {year} {1994})},\ \Eprint
  {https://arxiv.org/abs/hep-ph/9406363} {arXiv:hep-ph/9406363 [hep-ph]}
  \BibitemShut {NoStop}%
\bibitem [{\citenamefont {Chetyrkin}\ \emph
  {et~al.}(1995{\natexlab{a}})\citenamefont {Chetyrkin}, \citenamefont {Kuhn},\
  and\ \citenamefont {Steinhauser}}]{Chetyrkin:1995ix}%
  \BibitemOpen
  \bibfield  {author} {\bibinfo {author} {\bibfnamefont {K.}~\bibnamefont
  {Chetyrkin}}, \bibinfo {author} {\bibfnamefont {J.~H.}\ \bibnamefont
  {Kuhn}},\ and\ \bibinfo {author} {\bibfnamefont {M.}~\bibnamefont
  {Steinhauser}},\ }\bibfield  {title} {\bibinfo {title} {{Corrections of order
  ${\cal O}(G_F M_t^2 \alpha_s^2)$ to the $\rho$ parameter}},\ }\href
  {https://doi.org/10.1016/0370-2693(95)00380-4} {\bibfield  {journal}
  {\bibinfo  {journal} {Phys.Lett.}\ }\textbf {\bibinfo {volume} {B351}},\
  \bibinfo {pages} {331} (\bibinfo {year} {1995}{\natexlab{a}})},\ \Eprint
  {https://arxiv.org/abs/hep-ph/9502291} {arXiv:hep-ph/9502291 [hep-ph]}
  \BibitemShut {NoStop}%
\bibitem [{\citenamefont {Chetyrkin}\ \emph
  {et~al.}(1995{\natexlab{b}})\citenamefont {Chetyrkin}, \citenamefont {Kuhn},\
  and\ \citenamefont {Steinhauser}}]{Chetyrkin:1995js}%
  \BibitemOpen
  \bibfield  {author} {\bibinfo {author} {\bibfnamefont {K.}~\bibnamefont
  {Chetyrkin}}, \bibinfo {author} {\bibfnamefont {J.~H.}\ \bibnamefont
  {Kuhn}},\ and\ \bibinfo {author} {\bibfnamefont {M.}~\bibnamefont
  {Steinhauser}},\ }\bibfield  {title} {\bibinfo {title} {{QCD corrections from
  top quark to relations between electroweak parameters to order
  $\alpha_s^2$}},\ }\href {https://doi.org/10.1103/PhysRevLett.75.3394}
  {\bibfield  {journal} {\bibinfo  {journal} {Phys.Rev.Lett.}\ }\textbf
  {\bibinfo {volume} {75}},\ \bibinfo {pages} {3394} (\bibinfo {year}
  {1995}{\natexlab{b}})},\ \Eprint {https://arxiv.org/abs/hep-ph/9504413}
  {arXiv:hep-ph/9504413 [hep-ph]} \BibitemShut {NoStop}%
\bibitem [{\citenamefont {Degrassi}\ \emph {et~al.}(1996)\citenamefont
  {Degrassi}, \citenamefont {Gambino},\ and\ \citenamefont
  {Vicini}}]{Degrassi:1996mg}%
  \BibitemOpen
  \bibfield  {author} {\bibinfo {author} {\bibfnamefont {G.}~\bibnamefont
  {Degrassi}}, \bibinfo {author} {\bibfnamefont {P.}~\bibnamefont {Gambino}},\
  and\ \bibinfo {author} {\bibfnamefont {A.}~\bibnamefont {Vicini}},\
  }\bibfield  {title} {\bibinfo {title} {{Two-loop heavy top effects on the
  $m_Z$-$m_W$ interdependence}},\ }\href
  {https://doi.org/10.1016/0370-2693(96)00720-4} {\bibfield  {journal}
  {\bibinfo  {journal} {Phys.Lett.}\ }\textbf {\bibinfo {volume} {B383}},\
  \bibinfo {pages} {219} (\bibinfo {year} {1996})},\ \Eprint
  {https://arxiv.org/abs/hep-ph/9603374} {arXiv:hep-ph/9603374 [hep-ph]}
  \BibitemShut {NoStop}%
\bibitem [{\citenamefont {Degrassi}\ \emph {et~al.}(1997)\citenamefont
  {Degrassi}, \citenamefont {Gambino},\ and\ \citenamefont
  {Sirlin}}]{Degrassi:1996ps}%
  \BibitemOpen
  \bibfield  {author} {\bibinfo {author} {\bibfnamefont {G.}~\bibnamefont
  {Degrassi}}, \bibinfo {author} {\bibfnamefont {P.}~\bibnamefont {Gambino}},\
  and\ \bibinfo {author} {\bibfnamefont {A.}~\bibnamefont {Sirlin}},\
  }\bibfield  {title} {\bibinfo {title} {{Precise calculation of $M_W$,
  $\sin^2\hat{\theta}_W(M_Z)$, and $\sin^2\theta_{\rm eff}^{\rm lept}$}},\
  }\href {https://doi.org/10.1016/S0370-2693(96)01677-2} {\bibfield  {journal}
  {\bibinfo  {journal} {Phys.Lett.}\ }\textbf {\bibinfo {volume} {B394}},\
  \bibinfo {pages} {188} (\bibinfo {year} {1997})},\ \Eprint
  {https://arxiv.org/abs/hep-ph/9611363} {arXiv:hep-ph/9611363 [hep-ph]}
  \BibitemShut {NoStop}%
\bibitem [{\citenamefont {Degrassi}\ and\ \citenamefont
  {Gambino}(2000)}]{Degrassi:1999jd}%
  \BibitemOpen
  \bibfield  {author} {\bibinfo {author} {\bibfnamefont {G.}~\bibnamefont
  {Degrassi}}\ and\ \bibinfo {author} {\bibfnamefont {P.}~\bibnamefont
  {Gambino}},\ }\bibfield  {title} {\bibinfo {title} {{Two-loop heavy top
  corrections to the $Z^0$ boson partial widths}},\ }\href
  {https://doi.org/10.1016/S0550-3213(99)00729-4} {\bibfield  {journal}
  {\bibinfo  {journal} {Nucl.Phys.}\ }\textbf {\bibinfo {volume} {B567}},\
  \bibinfo {pages} {3} (\bibinfo {year} {2000})},\ \Eprint
  {https://arxiv.org/abs/hep-ph/9905472} {arXiv:hep-ph/9905472 [hep-ph]}
  \BibitemShut {NoStop}%
\bibitem [{\citenamefont {Freitas}\ \emph {et~al.}(2000)\citenamefont
  {Freitas}, \citenamefont {Hollik}, \citenamefont {Walter},\ and\
  \citenamefont {Weiglein}}]{Freitas:2000gg}%
  \BibitemOpen
  \bibfield  {author} {\bibinfo {author} {\bibfnamefont {A.}~\bibnamefont
  {Freitas}}, \bibinfo {author} {\bibfnamefont {W.}~\bibnamefont {Hollik}},
  \bibinfo {author} {\bibfnamefont {W.}~\bibnamefont {Walter}},\ and\ \bibinfo
  {author} {\bibfnamefont {G.}~\bibnamefont {Weiglein}},\ }\bibfield  {title}
  {\bibinfo {title} {{Complete fermionic two-loop results for the $M_W$-$M_Z$
  interdependence}},\ }\href {https://doi.org/10.1016/S0370-2693(00)01263-6}
  {\bibfield  {journal} {\bibinfo  {journal} {Phys.Lett.}\ }\textbf {\bibinfo
  {volume} {B495}},\ \bibinfo {pages} {338} (\bibinfo {year} {2000})},\ \Eprint
  {https://arxiv.org/abs/hep-ph/0007091} {arXiv:hep-ph/0007091 [hep-ph]}
  \BibitemShut {NoStop}%
\bibitem [{\citenamefont {van~der Bij}\ \emph {et~al.}(2001)\citenamefont
  {van~der Bij}, \citenamefont {Chetyrkin}, \citenamefont {Faisst},
  \citenamefont {Jikia},\ and\ \citenamefont
  {Seidensticker}}]{vanderBij:2000cg}%
  \BibitemOpen
  \bibfield  {author} {\bibinfo {author} {\bibfnamefont {J.}~\bibnamefont
  {van~der Bij}}, \bibinfo {author} {\bibfnamefont {K.}~\bibnamefont
  {Chetyrkin}}, \bibinfo {author} {\bibfnamefont {M.}~\bibnamefont {Faisst}},
  \bibinfo {author} {\bibfnamefont {G.}~\bibnamefont {Jikia}},\ and\ \bibinfo
  {author} {\bibfnamefont {T.}~\bibnamefont {Seidensticker}},\ }\bibfield
  {title} {\bibinfo {title} {{Three-loop leading top mass contributions to the
  $\rho$ parameter}},\ }\href {https://doi.org/10.1016/S0370-2693(01)00002-8}
  {\bibfield  {journal} {\bibinfo  {journal} {Phys.Lett.}\ }\textbf {\bibinfo
  {volume} {B498}},\ \bibinfo {pages} {156} (\bibinfo {year} {2001})},\ \Eprint
  {https://arxiv.org/abs/hep-ph/0011373} {arXiv:hep-ph/0011373 [hep-ph]}
  \BibitemShut {NoStop}%
\bibitem [{\citenamefont {Freitas}\ \emph {et~al.}(2002)\citenamefont
  {Freitas}, \citenamefont {Hollik}, \citenamefont {Walter},\ and\
  \citenamefont {Weiglein}}]{Freitas:2002ja}%
  \BibitemOpen
  \bibfield  {author} {\bibinfo {author} {\bibfnamefont {A.}~\bibnamefont
  {Freitas}}, \bibinfo {author} {\bibfnamefont {W.}~\bibnamefont {Hollik}},
  \bibinfo {author} {\bibfnamefont {W.}~\bibnamefont {Walter}},\ and\ \bibinfo
  {author} {\bibfnamefont {G.}~\bibnamefont {Weiglein}},\ }\bibfield  {title}
  {\bibinfo {title} {{Electroweak two-loop corrections to the $M_W$-$M_Z$ mass
  correlation in the standard model}},\ }\href
  {https://doi.org/10.1016/S0550-3213(02)00243-2} {\bibfield  {journal}
  {\bibinfo  {journal} {Nucl.Phys.}\ }\textbf {\bibinfo {volume} {B632}},\
  \bibinfo {pages} {189} (\bibinfo {year} {2002})},\ \Eprint
  {https://arxiv.org/abs/hep-ph/0202131} {arXiv:hep-ph/0202131 [hep-ph]}
  \BibitemShut {NoStop}%
\bibitem [{\citenamefont {Awramik}\ and\ \citenamefont
  {Czakon}(2002)}]{Awramik:2002wn}%
  \BibitemOpen
  \bibfield  {author} {\bibinfo {author} {\bibfnamefont {M.}~\bibnamefont
  {Awramik}}\ and\ \bibinfo {author} {\bibfnamefont {M.}~\bibnamefont
  {Czakon}},\ }\bibfield  {title} {\bibinfo {title} {{Complete two loop bosonic
  contributions to the muon lifetime in the standard model}},\ }\href
  {https://doi.org/10.1103/PhysRevLett.89.241801} {\bibfield  {journal}
  {\bibinfo  {journal} {Phys.Rev.Lett.}\ }\textbf {\bibinfo {volume} {89}},\
  \bibinfo {pages} {241801} (\bibinfo {year} {2002})},\ \Eprint
  {https://arxiv.org/abs/hep-ph/0208113} {arXiv:hep-ph/0208113 [hep-ph]}
  \BibitemShut {NoStop}%
\bibitem [{\citenamefont {Onishchenko}\ and\ \citenamefont
  {Veretin}(2003)}]{Onishchenko:2002ve}%
  \BibitemOpen
  \bibfield  {author} {\bibinfo {author} {\bibfnamefont {A.}~\bibnamefont
  {Onishchenko}}\ and\ \bibinfo {author} {\bibfnamefont {O.}~\bibnamefont
  {Veretin}},\ }\bibfield  {title} {\bibinfo {title} {{Two-loop bosonic
  electroweak corrections to the muon lifetime and $M_Z$-$M_W$
  interdependence}},\ }\href {https://doi.org/10.1016/S0370-2693(02)03004-6}
  {\bibfield  {journal} {\bibinfo  {journal} {Phys.Lett.}\ }\textbf {\bibinfo
  {volume} {B551}},\ \bibinfo {pages} {111} (\bibinfo {year} {2003})},\ \Eprint
  {https://arxiv.org/abs/hep-ph/0209010} {arXiv:hep-ph/0209010 [hep-ph]}
  \BibitemShut {NoStop}%
\bibitem [{\citenamefont {Awramik}\ \emph {et~al.}(2003)\citenamefont
  {Awramik}, \citenamefont {Czakon}, \citenamefont {Onishchenko},\ and\
  \citenamefont {Veretin}}]{Awramik:2002vu}%
  \BibitemOpen
  \bibfield  {author} {\bibinfo {author} {\bibfnamefont {M.}~\bibnamefont
  {Awramik}}, \bibinfo {author} {\bibfnamefont {M.}~\bibnamefont {Czakon}},
  \bibinfo {author} {\bibfnamefont {A.}~\bibnamefont {Onishchenko}},\ and\
  \bibinfo {author} {\bibfnamefont {O.}~\bibnamefont {Veretin}},\ }\bibfield
  {title} {\bibinfo {title} {{Bosonic corrections to $\Delta r$ at the two loop
  level}},\ }\href {https://doi.org/10.1103/PhysRevD.68.053004} {\bibfield
  {journal} {\bibinfo  {journal} {Phys.Rev.}\ }\textbf {\bibinfo {volume}
  {D68}},\ \bibinfo {pages} {053004} (\bibinfo {year} {2003})},\ \Eprint
  {https://arxiv.org/abs/hep-ph/0209084} {arXiv:hep-ph/0209084 [hep-ph]}
  \BibitemShut {NoStop}%
\bibitem [{\citenamefont {Awramik}\ and\ \citenamefont
  {Czakon}(2003{\natexlab{a}})}]{Awramik:2002wv}%
  \BibitemOpen
  \bibfield  {author} {\bibinfo {author} {\bibfnamefont {M.}~\bibnamefont
  {Awramik}}\ and\ \bibinfo {author} {\bibfnamefont {M.}~\bibnamefont
  {Czakon}},\ }\bibfield  {title} {\bibinfo {title} {{Two loop electroweak
  bosonic corrections to the muon decay lifetime}},\ }\href
  {https://doi.org/10.1016/S0920-5632(03)80177-9} {\bibfield  {journal}
  {\bibinfo  {journal} {Nucl.Phys.Proc.Suppl.}\ }\textbf {\bibinfo {volume}
  {116}},\ \bibinfo {pages} {238} (\bibinfo {year} {2003}{\natexlab{a}})},\
  \Eprint {https://arxiv.org/abs/hep-ph/0211041} {arXiv:hep-ph/0211041
  [hep-ph]} \BibitemShut {NoStop}%
\bibitem [{\citenamefont {Awramik}\ and\ \citenamefont
  {Czakon}(2003{\natexlab{b}})}]{Awramik:2003ee}%
  \BibitemOpen
  \bibfield  {author} {\bibinfo {author} {\bibfnamefont {M.}~\bibnamefont
  {Awramik}}\ and\ \bibinfo {author} {\bibfnamefont {M.}~\bibnamefont
  {Czakon}},\ }\bibfield  {title} {\bibinfo {title} {{Complete two loop
  electroweak contributions to the muon lifetime in the standard model}},\
  }\href {https://doi.org/10.1016/j.physletb.2003.06.007} {\bibfield  {journal}
  {\bibinfo  {journal} {Phys.Lett.}\ }\textbf {\bibinfo {volume} {B568}},\
  \bibinfo {pages} {48} (\bibinfo {year} {2003}{\natexlab{b}})},\ \Eprint
  {https://arxiv.org/abs/hep-ph/0305248} {arXiv:hep-ph/0305248 [hep-ph]}
  \BibitemShut {NoStop}%
\bibitem [{\citenamefont {Awramik}\ \emph {et~al.}(2004)\citenamefont
  {Awramik}, \citenamefont {Czakon}, \citenamefont {Freitas},\ and\
  \citenamefont {Weiglein}}]{Awramik:2003rn}%
  \BibitemOpen
  \bibfield  {author} {\bibinfo {author} {\bibfnamefont {M.}~\bibnamefont
  {Awramik}}, \bibinfo {author} {\bibfnamefont {M.}~\bibnamefont {Czakon}},
  \bibinfo {author} {\bibfnamefont {A.}~\bibnamefont {Freitas}},\ and\ \bibinfo
  {author} {\bibfnamefont {G.}~\bibnamefont {Weiglein}},\ }\bibfield  {title}
  {\bibinfo {title} {{Precise prediction for the $W$ boson mass in the standard
  model}},\ }\href {https://doi.org/10.1103/PhysRevD.69.053006} {\bibfield
  {journal} {\bibinfo  {journal} {Phys.Rev.}\ }\textbf {\bibinfo {volume}
  {D69}},\ \bibinfo {pages} {053006} (\bibinfo {year} {2004})},\ \Eprint
  {https://arxiv.org/abs/hep-ph/0311148} {arXiv:hep-ph/0311148 [hep-ph]}
  \BibitemShut {NoStop}%
\bibitem [{\citenamefont {Faisst}\ \emph {et~al.}(2003)\citenamefont {Faisst},
  \citenamefont {Kuhn}, \citenamefont {Seidensticker},\ and\ \citenamefont
  {Veretin}}]{Faisst:2003px}%
  \BibitemOpen
  \bibfield  {author} {\bibinfo {author} {\bibfnamefont {M.}~\bibnamefont
  {Faisst}}, \bibinfo {author} {\bibfnamefont {J.~H.}\ \bibnamefont {Kuhn}},
  \bibinfo {author} {\bibfnamefont {T.}~\bibnamefont {Seidensticker}},\ and\
  \bibinfo {author} {\bibfnamefont {O.}~\bibnamefont {Veretin}},\ }\bibfield
  {title} {\bibinfo {title} {{Three loop top quark contributions to the rho
  parameter}},\ }\href {https://doi.org/10.1016/S0550-3213(03)00450-4}
  {\bibfield  {journal} {\bibinfo  {journal} {Nucl. Phys.}\ }\textbf {\bibinfo
  {volume} {B665}},\ \bibinfo {pages} {649} (\bibinfo {year} {2003})},\ \Eprint
  {https://arxiv.org/abs/hep-ph/0302275} {arXiv:hep-ph/0302275 [hep-ph]}
  \BibitemShut {NoStop}%
\bibitem [{\citenamefont {Dubovyk}\ \emph {et~al.}(2016)\citenamefont
  {Dubovyk}, \citenamefont {Freitas}, \citenamefont {Gluza}, \citenamefont
  {Riemann},\ and\ \citenamefont {Usovitsch}}]{Dubovyk:2016aqv}%
  \BibitemOpen
  \bibfield  {author} {\bibinfo {author} {\bibfnamefont {I.}~\bibnamefont
  {Dubovyk}}, \bibinfo {author} {\bibfnamefont {A.}~\bibnamefont {Freitas}},
  \bibinfo {author} {\bibfnamefont {J.}~\bibnamefont {Gluza}}, \bibinfo
  {author} {\bibfnamefont {T.}~\bibnamefont {Riemann}},\ and\ \bibinfo {author}
  {\bibfnamefont {J.}~\bibnamefont {Usovitsch}},\ }\bibfield  {title} {\bibinfo
  {title} {{The two-loop electroweak bosonic corrections to
  $\sin^2\theta^\textrm{b}_\textrm{eff}$}},\ }\href
  {https://doi.org/10.1016/j.physletb.2016.09.012} {\bibfield  {journal}
  {\bibinfo  {journal} {Phys. Lett. B}\ }\textbf {\bibinfo {volume} {762}},\
  \bibinfo {pages} {184} (\bibinfo {year} {2016})},\ \Eprint
  {https://arxiv.org/abs/1607.08375} {arXiv:1607.08375 [hep-ph]} \BibitemShut
  {NoStop}%
\bibitem [{\citenamefont {Dubovyk}\ \emph {et~al.}(2018)\citenamefont
  {Dubovyk}, \citenamefont {Freitas}, \citenamefont {Gluza}, \citenamefont
  {Riemann},\ and\ \citenamefont {Usovitsch}}]{Dubovyk:2018rlg}%
  \BibitemOpen
  \bibfield  {author} {\bibinfo {author} {\bibfnamefont {I.}~\bibnamefont
  {Dubovyk}}, \bibinfo {author} {\bibfnamefont {A.}~\bibnamefont {Freitas}},
  \bibinfo {author} {\bibfnamefont {J.}~\bibnamefont {Gluza}}, \bibinfo
  {author} {\bibfnamefont {T.}~\bibnamefont {Riemann}},\ and\ \bibinfo {author}
  {\bibfnamefont {J.}~\bibnamefont {Usovitsch}},\ }\bibfield  {title} {\bibinfo
  {title} {{Complete electroweak two-loop corrections to Z boson production and
  decay}},\ }\href {https://doi.org/10.1016/j.physletb.2018.06.037} {\bibfield
  {journal} {\bibinfo  {journal} {Phys. Lett. B}\ }\textbf {\bibinfo {volume}
  {783}},\ \bibinfo {pages} {86} (\bibinfo {year} {2018})},\ \Eprint
  {https://arxiv.org/abs/1804.10236} {arXiv:1804.10236 [hep-ph]} \BibitemShut
  {NoStop}%
\bibitem [{\citenamefont {Schael}\ \emph {et~al.}(2006)\citenamefont {Schael}
  \emph {et~al.}}]{ALEPH:2005ab}%
  \BibitemOpen
  \bibfield  {author} {\bibinfo {author} {\bibfnamefont {S.}~\bibnamefont
  {Schael}} \emph {et~al.} (\bibinfo {collaboration} {ALEPH, DELPHI, L3, OPAL,
  SLD, LEP Electroweak Working Group, SLD Electroweak Group, SLD Heavy Flavour
  Group}),\ }\bibfield  {title} {\bibinfo {title} {{Precision electroweak
  measurements on the $Z$ resonance}},\ }\href
  {https://doi.org/10.1016/j.physrep.2005.12.006} {\bibfield  {journal}
  {\bibinfo  {journal} {Phys. Rept.}\ }\textbf {\bibinfo {volume} {427}},\
  \bibinfo {pages} {257} (\bibinfo {year} {2006})},\ \Eprint
  {https://arxiv.org/abs/hep-ex/0509008} {arXiv:hep-ex/0509008} \BibitemShut
  {NoStop}%
\bibitem [{\citenamefont {{The ALEPH, DELPHI, L3, OPAL Collaborations, the LEP
  Electroweak Working Group}}(2013)}]{LEP-2}%
  \BibitemOpen
  \bibfield  {author} {\bibinfo {author} {\bibnamefont {{The ALEPH, DELPHI, L3,
  OPAL Collaborations, the LEP Electroweak Working Group}}},\ }\bibfield
  {title} {\bibinfo {title} {{Electroweak Measurements in Electron-Positron
  Collisions at W-Boson-Pair Energies at LEP}},\ }\href@noop {} {\bibfield
  {journal} {\bibinfo  {journal} {Phys. Rept.}\ }\textbf {\bibinfo {volume}
  {532}},\ \bibinfo {pages} {119} (\bibinfo {year} {2013})},\ \Eprint
  {https://arxiv.org/abs/1302.3415} {arXiv:1302.3415 [hep-ex]} \BibitemShut
  {NoStop}%
\bibitem [{\citenamefont {{ALEPH, CDF, D0, DELPHI, L3, OPAL, SLD
  Collaborations, LEP Electroweak Working Group, Tevatron Electroweak Working
  Group, and SLD Electroweak and Heavy Flavour Groups}}(2010)}]{ALEPH:2010aa}%
  \BibitemOpen
  \bibfield  {author} {\bibinfo {author} {\bibnamefont {{ALEPH, CDF, D0,
  DELPHI, L3, OPAL, SLD Collaborations, LEP Electroweak Working Group, Tevatron
  Electroweak Working Group, and SLD Electroweak and Heavy Flavour Groups}}},\
  }\bibfield  {title} {\bibinfo {title} {{Precision electroweak measurements
  and constraints on the standard model}},\ }\Eprint
  {https://arxiv.org/abs/1012.2367} {arXiv:1012.2367 [hep-ex]}  (\bibinfo
  {year} {2010})\BibitemShut {NoStop}%
\bibitem [{CDF(2016{\natexlab{a}})}]{CDF:2016cry}%
  \BibitemOpen
  \bibfield  {title} {\bibinfo {title} {{Combination of the CDF and D0
  Effective Leptonic Electroweak Mixing Angles}}\ }(\bibinfo {year}
  {2016})\BibitemShut {NoStop}%
\bibitem [{CDF(2016{\natexlab{b}})}]{CDF:2016vzt}%
  \BibitemOpen
  \bibfield  {title} {\bibinfo {title} {{Combination of CDF and D0 results on
  the mass of the top quark using up $9.7\:{\rm fb}^{-1}$ at the Tevatron}},\
  }\Eprint {https://arxiv.org/abs/1608.01881} {arXiv:1608.01881 [hep-ex]}
  (\bibinfo {year} {2016}{\natexlab{b}})\BibitemShut {NoStop}%
\bibitem [{\citenamefont {Khachatryan}\ \emph {et~al.}(2016)\citenamefont
  {Khachatryan} \emph {et~al.}}]{Khachatryan:2015hba}%
  \BibitemOpen
  \bibfield  {author} {\bibinfo {author} {\bibfnamefont {V.}~\bibnamefont
  {Khachatryan}} \emph {et~al.} (\bibinfo {collaboration} {CMS}),\ }\bibfield
  {title} {\bibinfo {title} {{Measurement of the top quark mass using
  proton-proton data at ${\sqrt{(s)}}$ = 7 and 8 TeV}},\ }\href
  {https://doi.org/10.1103/PhysRevD.93.072004} {\bibfield  {journal} {\bibinfo
  {journal} {Phys. Rev. D}\ }\textbf {\bibinfo {volume} {93}},\ \bibinfo
  {pages} {072004} (\bibinfo {year} {2016})},\ \Eprint
  {https://arxiv.org/abs/1509.04044} {arXiv:1509.04044 [hep-ex]} \BibitemShut
  {NoStop}%
\bibitem [{\citenamefont {Aaboud}\ \emph {et~al.}(2019)\citenamefont {Aaboud}
  \emph {et~al.}}]{Aaboud:2018zbu}%
  \BibitemOpen
  \bibfield  {author} {\bibinfo {author} {\bibfnamefont {M.}~\bibnamefont
  {Aaboud}} \emph {et~al.} (\bibinfo {collaboration} {ATLAS}),\ }\bibfield
  {title} {\bibinfo {title} {{Measurement of the top quark mass in the
  $t\bar{t}\rightarrow $ lepton+jets channel from $\sqrt{s}=8$ TeV ATLAS data
  and combination with previous results}},\ }\href
  {https://doi.org/10.1140/epjc/s10052-019-6757-9} {\bibfield  {journal}
  {\bibinfo  {journal} {Eur. Phys. J. C}\ }\textbf {\bibinfo {volume} {79}},\
  \bibinfo {pages} {290} (\bibinfo {year} {2019})},\ \Eprint
  {https://arxiv.org/abs/1810.01772} {arXiv:1810.01772 [hep-ex]} \BibitemShut
  {NoStop}%
\bibitem [{\citenamefont {Sirunyan}\ \emph
  {et~al.}(2018{\natexlab{a}})\citenamefont {Sirunyan} \emph
  {et~al.}}]{Sirunyan:2018gqx}%
  \BibitemOpen
  \bibfield  {author} {\bibinfo {author} {\bibfnamefont {A.~M.}\ \bibnamefont
  {Sirunyan}} \emph {et~al.} (\bibinfo {collaboration} {CMS}),\ }\bibfield
  {title} {\bibinfo {title} {{Measurement of the top quark mass with
  lepton+jets final states using $\mathrm {p}$ $\mathrm {p}$ collisions at
  $\sqrt{s}=13\,\text {TeV} $}},\ }\href
  {https://doi.org/10.1140/epjc/s10052-018-6332-9} {\bibfield  {journal}
  {\bibinfo  {journal} {Eur. Phys. J. C}\ }\textbf {\bibinfo {volume} {78}},\
  \bibinfo {pages} {891} (\bibinfo {year} {2018}{\natexlab{a}})},\ \Eprint
  {https://arxiv.org/abs/1805.01428} {arXiv:1805.01428 [hep-ex]} \BibitemShut
  {NoStop}%
\bibitem [{\citenamefont {Sirunyan}\ \emph
  {et~al.}(2019{\natexlab{a}})\citenamefont {Sirunyan} \emph
  {et~al.}}]{Sirunyan:2018goh}%
  \BibitemOpen
  \bibfield  {author} {\bibinfo {author} {\bibfnamefont {A.~M.}\ \bibnamefont
  {Sirunyan}} \emph {et~al.} (\bibinfo {collaboration} {CMS}),\ }\bibfield
  {title} {\bibinfo {title} {{Measurement of the
  $\mathrm{t}\overline{\mathrm{t}}$ production cross section, the top quark
  mass, and the strong coupling constant using dilepton events in pp collisions
  at $\sqrt{s} =$ 13 TeV}},\ }\href
  {https://doi.org/10.1140/epjc/s10052-019-6863-8} {\bibfield  {journal}
  {\bibinfo  {journal} {Eur. Phys. J. C}\ }\textbf {\bibinfo {volume} {79}},\
  \bibinfo {pages} {368} (\bibinfo {year} {2019}{\natexlab{a}})},\ \Eprint
  {https://arxiv.org/abs/1812.10505} {arXiv:1812.10505 [hep-ex]} \BibitemShut
  {NoStop}%
\bibitem [{\citenamefont {Sirunyan}\ \emph
  {et~al.}(2019{\natexlab{b}})\citenamefont {Sirunyan} \emph
  {et~al.}}]{Sirunyan:2018mlv}%
  \BibitemOpen
  \bibfield  {author} {\bibinfo {author} {\bibfnamefont {A.~M.}\ \bibnamefont
  {Sirunyan}} \emph {et~al.} (\bibinfo {collaboration} {CMS}),\ }\bibfield
  {title} {\bibinfo {title} {{Measurement of the top quark mass in the all-jets
  final state at $\sqrt{s} =$ 13 TeV and combination with the lepton+jets
  channel}},\ }\href {https://doi.org/10.1140/epjc/s10052-019-6788-2}
  {\bibfield  {journal} {\bibinfo  {journal} {Eur. Phys. J. C}\ }\textbf
  {\bibinfo {volume} {79}},\ \bibinfo {pages} {313} (\bibinfo {year}
  {2019}{\natexlab{b}})},\ \Eprint {https://arxiv.org/abs/1812.10534}
  {arXiv:1812.10534 [hep-ex]} \BibitemShut {NoStop}%
\bibitem [{ATL(2019)}]{ATLAS:2019ezb}%
  \BibitemOpen
  \href@noop {} {\bibinfo {title} {{Measurement of the top quark mass using a
  leptonic invariant mass in pp collisions at $sqrt{s}$ = 13 TeV with the ATLAS
  detector}}} (\bibinfo {year} {2019})\BibitemShut {NoStop}%
\bibitem [{\citenamefont {Aaboud}\ \emph
  {et~al.}(2018{\natexlab{a}})\citenamefont {Aaboud} \emph
  {et~al.}}]{Aaboud:2017svj}%
  \BibitemOpen
  \bibfield  {author} {\bibinfo {author} {\bibfnamefont {M.}~\bibnamefont
  {Aaboud}} \emph {et~al.} (\bibinfo {collaboration} {ATLAS}),\ }\bibfield
  {title} {\bibinfo {title} {{Measurement of the $W$-boson mass in pp
  collisions at $\sqrt{s}=7$ TeV with the ATLAS detector}},\ }\href
  {https://doi.org/10.1140/epjc/s10052-017-5475-4} {\bibfield  {journal}
  {\bibinfo  {journal} {Eur. Phys. J. C}\ }\textbf {\bibinfo {volume} {78}},\
  \bibinfo {pages} {110} (\bibinfo {year} {2018}{\natexlab{a}})},\ \bibinfo
  {note} {[Erratum: Eur.Phys.J.C 78, 898 (2018)]},\ \Eprint
  {https://arxiv.org/abs/1701.07240} {arXiv:1701.07240 [hep-ex]} \BibitemShut
  {NoStop}%
\bibitem [{\citenamefont {Aad}\ \emph {et~al.}(2015{\natexlab{a}})\citenamefont
  {Aad} \emph {et~al.}}]{Aad:2015uau}%
  \BibitemOpen
  \bibfield  {author} {\bibinfo {author} {\bibfnamefont {G.}~\bibnamefont
  {Aad}} \emph {et~al.} (\bibinfo {collaboration} {ATLAS}),\ }\bibfield
  {title} {\bibinfo {title} {{Measurement of the forward-backward asymmetry of
  electron and muon pair-production in $pp$ collisions at $\sqrt{s}$ = 7 TeV
  with the ATLAS detector}},\ }\href {https://doi.org/10.1007/JHEP09(2015)049}
  {\bibfield  {journal} {\bibinfo  {journal} {JHEP}\ }\textbf {\bibinfo
  {volume} {09}},\ \bibinfo {pages} {049}},\ \Eprint
  {https://arxiv.org/abs/1503.03709} {arXiv:1503.03709 [hep-ex]} \BibitemShut
  {NoStop}%
\bibitem [{ATL(2018)}]{ATLAS:2018gqq}%
  \BibitemOpen
  \href@noop {} {\bibinfo {title} {{Measurement of the effective leptonic weak
  mixing angle using electron and muon pairs from $Z$-boson decay in the ATLAS
  experiment at $\sqrt s = 8$ TeV}}} (\bibinfo {year} {2018})\BibitemShut
  {NoStop}%
\bibitem [{\citenamefont {Sirunyan}\ \emph
  {et~al.}(2018{\natexlab{b}})\citenamefont {Sirunyan} \emph
  {et~al.}}]{Sirunyan:2018swq}%
  \BibitemOpen
  \bibfield  {author} {\bibinfo {author} {\bibfnamefont {A.~M.}\ \bibnamefont
  {Sirunyan}} \emph {et~al.} (\bibinfo {collaboration} {CMS}),\ }\bibfield
  {title} {\bibinfo {title} {{Measurement of the weak mixing angle using the
  forward-backward asymmetry of Drell-Yan events in pp collisions at 8 TeV}},\
  }\href {https://doi.org/10.1140/epjc/s10052-018-6148-7} {\bibfield  {journal}
  {\bibinfo  {journal} {Eur. Phys. J. C}\ }\textbf {\bibinfo {volume} {78}},\
  \bibinfo {pages} {701} (\bibinfo {year} {2018}{\natexlab{b}})},\ \Eprint
  {https://arxiv.org/abs/1806.00863} {arXiv:1806.00863 [hep-ex]} \BibitemShut
  {NoStop}%
\bibitem [{\citenamefont {Aaij}\ \emph {et~al.}(2015)\citenamefont {Aaij} \emph
  {et~al.}}]{Aaij:2015lka}%
  \BibitemOpen
  \bibfield  {author} {\bibinfo {author} {\bibfnamefont {R.}~\bibnamefont
  {Aaij}} \emph {et~al.} (\bibinfo {collaboration} {LHCb}),\ }\bibfield
  {title} {\bibinfo {title} {{Measurement of the forward-backward asymmetry in
  $Z/\gamma^{\ast} \rightarrow \mu^{+}\mu^{-}$ decays and determination of the
  effective weak mixing angle}},\ }\href
  {https://doi.org/10.1007/JHEP11(2015)190} {\bibfield  {journal} {\bibinfo
  {journal} {JHEP}\ }\textbf {\bibinfo {volume} {11}},\ \bibinfo {pages}
  {190}},\ \Eprint {https://arxiv.org/abs/1509.07645} {arXiv:1509.07645
  [hep-ex]} \BibitemShut {NoStop}%
\bibitem [{\citenamefont {Aad}\ \emph {et~al.}(2015{\natexlab{b}})\citenamefont
  {Aad} \emph {et~al.}}]{Aad:2015zhl}%
  \BibitemOpen
  \bibfield  {author} {\bibinfo {author} {\bibfnamefont {G.}~\bibnamefont
  {Aad}} \emph {et~al.} (\bibinfo {collaboration} {ATLAS, CMS}),\ }\bibfield
  {title} {\bibinfo {title} {{Combined Measurement of the Higgs Boson Mass in
  $pp$ Collisions at $\sqrt{s}=7$ and 8 TeV with the ATLAS and CMS
  Experiments}},\ }\href {https://doi.org/10.1103/PhysRevLett.114.191803}
  {\bibfield  {journal} {\bibinfo  {journal} {Phys. Rev. Lett.}\ }\textbf
  {\bibinfo {volume} {114}},\ \bibinfo {pages} {191803} (\bibinfo {year}
  {2015}{\natexlab{b}})},\ \Eprint {https://arxiv.org/abs/1503.07589}
  {arXiv:1503.07589 [hep-ex]} \BibitemShut {NoStop}%
\bibitem [{\citenamefont {de~Blas}\ \emph {et~al.}(2020)\citenamefont {de~Blas}
  \emph {et~al.}}]{deBlas:2019okz}%
  \BibitemOpen
  \bibfield  {author} {\bibinfo {author} {\bibfnamefont {J.}~\bibnamefont
  {de~Blas}} \emph {et~al.},\ }\bibfield  {title} {\bibinfo {title}
  {{$\texttt{HEPfit}$: a code for the combination of indirect and direct
  constraints on high energy physics models}},\ }\href
  {https://doi.org/10.1140/epjc/s10052-020-7904-z} {\bibfield  {journal}
  {\bibinfo  {journal} {Eur. Phys. J. C}\ }\textbf {\bibinfo {volume} {80}},\
  \bibinfo {pages} {456} (\bibinfo {year} {2020})},\ \Eprint
  {https://arxiv.org/abs/1910.14012} {arXiv:1910.14012 [hep-ph]} \BibitemShut
  {NoStop}%
\bibitem [{HEP()}]{HEPfit}%
  \BibitemOpen
  \href@noop {} {\bibinfo {title} {{\texttt{HEPfit}: a tool to combine indirect
  and direct constraints on High Energy Physics}}},\ \bibinfo {howpublished}
  {\url{http://hepfit.roma1.infn.it/}}\BibitemShut {NoStop}%
\bibitem [{\citenamefont {Caldwell}\ \emph {et~al.}(2009)\citenamefont
  {Caldwell}, \citenamefont {Kollar},\ and\ \citenamefont
  {Kroninger}}]{Caldwell:2008fw}%
  \BibitemOpen
  \bibfield  {author} {\bibinfo {author} {\bibfnamefont {A.}~\bibnamefont
  {Caldwell}}, \bibinfo {author} {\bibfnamefont {D.}~\bibnamefont {Kollar}},\
  and\ \bibinfo {author} {\bibfnamefont {K.}~\bibnamefont {Kroninger}},\
  }\bibfield  {title} {\bibinfo {title} {{BAT: The Bayesian Analysis
  Toolkit}},\ }\href {https://doi.org/10.1016/j.cpc.2009.06.026} {\bibfield
  {journal} {\bibinfo  {journal} {Comput.Phys.Commun.}\ }\textbf {\bibinfo
  {volume} {180}},\ \bibinfo {pages} {2197} (\bibinfo {year} {2009})},\ \Eprint
  {https://arxiv.org/abs/0808.2552} {arXiv:0808.2552 [physics.data-an]}
  \BibitemShut {NoStop}%
\bibitem [{\citenamefont {Ciuchini}\ \emph {et~al.}(2013)\citenamefont
  {Ciuchini}, \citenamefont {Franco}, \citenamefont {Mishima},\ and\
  \citenamefont {Silvestrini}}]{Ciuchini:2013pca}%
  \BibitemOpen
  \bibfield  {author} {\bibinfo {author} {\bibfnamefont {M.}~\bibnamefont
  {Ciuchini}}, \bibinfo {author} {\bibfnamefont {E.}~\bibnamefont {Franco}},
  \bibinfo {author} {\bibfnamefont {S.}~\bibnamefont {Mishima}},\ and\ \bibinfo
  {author} {\bibfnamefont {L.}~\bibnamefont {Silvestrini}},\ }\bibfield
  {title} {\bibinfo {title} {{Electroweak Precision Observables, New Physics
  and the Nature of a 126 GeV Higgs Boson}},\ }\href
  {https://doi.org/10.1007/JHEP08(2013)106} {\bibfield  {journal} {\bibinfo
  {journal} {JHEP}\ }\textbf {\bibinfo {volume} {08}},\ \bibinfo {pages}
  {106}},\ \Eprint {https://arxiv.org/abs/1306.4644} {arXiv:1306.4644 [hep-ph]}
  \BibitemShut {NoStop}%
\bibitem [{\citenamefont {de~Blas}\ \emph {et~al.}(2016)\citenamefont
  {de~Blas}, \citenamefont {Ciuchini}, \citenamefont {Franco}, \citenamefont
  {Mishima}, \citenamefont {Pierini}, \citenamefont {Reina},\ and\
  \citenamefont {Silvestrini}}]{deBlas:2016ojx}%
  \BibitemOpen
  \bibfield  {author} {\bibinfo {author} {\bibfnamefont {J.}~\bibnamefont
  {de~Blas}}, \bibinfo {author} {\bibfnamefont {M.}~\bibnamefont {Ciuchini}},
  \bibinfo {author} {\bibfnamefont {E.}~\bibnamefont {Franco}}, \bibinfo
  {author} {\bibfnamefont {S.}~\bibnamefont {Mishima}}, \bibinfo {author}
  {\bibfnamefont {M.}~\bibnamefont {Pierini}}, \bibinfo {author} {\bibfnamefont
  {L.}~\bibnamefont {Reina}},\ and\ \bibinfo {author} {\bibfnamefont
  {L.}~\bibnamefont {Silvestrini}},\ }\bibfield  {title} {\bibinfo {title}
  {{Electroweak precision observables and Higgs-boson signal strengths in the
  Standard Model and beyond: present and future}},\ }\href
  {https://doi.org/10.1007/JHEP12(2016)135} {\bibfield  {journal} {\bibinfo
  {journal} {JHEP}\ }\textbf {\bibinfo {volume} {12}},\ \bibinfo {pages}
  {135}},\ \Eprint {https://arxiv.org/abs/1608.01509} {arXiv:1608.01509
  [hep-ph]} \BibitemShut {NoStop}%
\bibitem [{\citenamefont {Chen}\ and\ \citenamefont
  {Freitas}(2020)}]{Chen:2020xzx}%
  \BibitemOpen
  \bibfield  {author} {\bibinfo {author} {\bibfnamefont {L.}~\bibnamefont
  {Chen}}\ and\ \bibinfo {author} {\bibfnamefont {A.}~\bibnamefont {Freitas}},\
  }\bibfield  {title} {\bibinfo {title} {{Leading fermionic three-loop
  corrections to electroweak precision observables}},\ }\href
  {https://doi.org/10.1007/JHEP07(2020)210} {\bibfield  {journal} {\bibinfo
  {journal} {JHEP}\ }\textbf {\bibinfo {volume} {07}},\ \bibinfo {pages}
  {210}},\ \Eprint {https://arxiv.org/abs/2002.05845} {arXiv:2002.05845
  [hep-ph]} \BibitemShut {NoStop}%
\bibitem [{\citenamefont {Chen}\ and\ \citenamefont
  {Freitas}(2021)}]{Chen:2020xot}%
  \BibitemOpen
  \bibfield  {author} {\bibinfo {author} {\bibfnamefont {L.}~\bibnamefont
  {Chen}}\ and\ \bibinfo {author} {\bibfnamefont {A.}~\bibnamefont {Freitas}},\
  }\bibfield  {title} {\bibinfo {title} {{Mixed EW-QCD leading fermionic
  three-loop corrections at $\mathcal{O}(\alpha_s\alpha^2)$ to electroweak
  precision observables}},\ }\href {https://doi.org/10.1007/JHEP03(2021)215}
  {\bibfield  {journal} {\bibinfo  {journal} {JHEP}\ }\textbf {\bibinfo
  {volume} {03}},\ \bibinfo {pages} {215}},\ \Eprint
  {https://arxiv.org/abs/2012.08605} {arXiv:2012.08605 [hep-ph]} \BibitemShut
  {NoStop}%
\bibitem [{\citenamefont {Tanabashi}\ \emph {et~al.}(2018)\citenamefont
  {Tanabashi} \emph {et~al.}}]{Tanabashi:2018oca}%
  \BibitemOpen
  \bibfield  {author} {\bibinfo {author} {\bibfnamefont {M.}~\bibnamefont
  {Tanabashi}} \emph {et~al.} (\bibinfo {collaboration} {Particle Data
  Group}),\ }\bibfield  {title} {\bibinfo {title} {{Review of Particle
  Physics}},\ }\href {https://doi.org/10.1103/PhysRevD.98.030001} {\bibfield
  {journal} {\bibinfo  {journal} {Phys. Rev.}\ }\textbf {\bibinfo {volume}
  {D98}},\ \bibinfo {pages} {030001} (\bibinfo {year} {2018})}\BibitemShut
  {NoStop}%
\bibitem [{\citenamefont {Erler}\ and\ \citenamefont
  {Schott}(2019)}]{Erler:2019hds}%
  \BibitemOpen
  \bibfield  {author} {\bibinfo {author} {\bibfnamefont {J.}~\bibnamefont
  {Erler}}\ and\ \bibinfo {author} {\bibfnamefont {M.}~\bibnamefont {Schott}},\
  }\bibfield  {title} {\bibinfo {title} {{Electroweak Precision Tests of the
  Standard Model after the Discovery of the Higgs Boson}},\ }\href
  {https://doi.org/10.1016/j.ppnp.2019.02.007} {\bibfield  {journal} {\bibinfo
  {journal} {Prog. Part. Nucl. Phys.}\ }\textbf {\bibinfo {volume} {106}},\
  \bibinfo {pages} {68} (\bibinfo {year} {2019})},\ \Eprint
  {https://arxiv.org/abs/1902.05142} {arXiv:1902.05142 [hep-ph]} \BibitemShut
  {NoStop}%
\bibitem [{\citenamefont {Zyla}\ \emph {et~al.}(2020)\citenamefont {Zyla} \emph
  {et~al.}}]{Zyla:2020zbs}%
  \BibitemOpen
  \bibfield  {author} {\bibinfo {author} {\bibfnamefont {P.~A.}\ \bibnamefont
  {Zyla}} \emph {et~al.} (\bibinfo {collaboration} {Particle Data Group}),\
  }\bibfield  {title} {\bibinfo {title} {{Review of Particle Physics}},\ }\href
  {https://doi.org/10.1093/ptep/ptaa104} {\bibfield  {journal} {\bibinfo
  {journal} {PTEP}\ }\textbf {\bibinfo {volume} {2020}},\ \bibinfo {pages}
  {083C01} (\bibinfo {year} {2020})}\BibitemShut {NoStop}%
\bibitem [{\citenamefont {Borsanyi}\ \emph {et~al.}(2018)\citenamefont
  {Borsanyi} \emph {et~al.}}]{Borsanyi:2017zdw}%
  \BibitemOpen
  \bibfield  {author} {\bibinfo {author} {\bibfnamefont {S.}~\bibnamefont
  {Borsanyi}} \emph {et~al.} (\bibinfo {collaboration}
  {Budapest-Marseille-Wuppertal}),\ }\bibfield  {title} {\bibinfo {title}
  {{Hadronic vacuum polarization contribution to the anomalous magnetic moments
  of leptons from first principles}},\ }\href
  {https://doi.org/10.1103/PhysRevLett.121.022002} {\bibfield  {journal}
  {\bibinfo  {journal} {Phys. Rev. Lett.}\ }\textbf {\bibinfo {volume} {121}},\
  \bibinfo {pages} {022002} (\bibinfo {year} {2018})},\ \Eprint
  {https://arxiv.org/abs/1711.04980} {arXiv:1711.04980 [hep-lat]} \BibitemShut
  {NoStop}%
\bibitem [{\citenamefont {Colquhoun}\ \emph {et~al.}(2015)\citenamefont
  {Colquhoun}, \citenamefont {Dowdall}, \citenamefont {Davies}, \citenamefont
  {Hornbostel},\ and\ \citenamefont {Lepage}}]{Colquhoun:2014ica}%
  \BibitemOpen
  \bibfield  {author} {\bibinfo {author} {\bibfnamefont {B.}~\bibnamefont
  {Colquhoun}}, \bibinfo {author} {\bibfnamefont {R.~J.}\ \bibnamefont
  {Dowdall}}, \bibinfo {author} {\bibfnamefont {C.~T.~H.}\ \bibnamefont
  {Davies}}, \bibinfo {author} {\bibfnamefont {K.}~\bibnamefont {Hornbostel}},\
  and\ \bibinfo {author} {\bibfnamefont {G.~P.}\ \bibnamefont {Lepage}},\
  }\bibfield  {title} {\bibinfo {title} {{$\Upsilon$ and $\Upsilon^{\prime}$
  Leptonic Widths, $a_{\mu}^b$ and $m_b$ from full lattice QCD}},\ }\href
  {https://doi.org/10.1103/PhysRevD.91.074514} {\bibfield  {journal} {\bibinfo
  {journal} {Phys. Rev. D}\ }\textbf {\bibinfo {volume} {91}},\ \bibinfo
  {pages} {074514} (\bibinfo {year} {2015})},\ \Eprint
  {https://arxiv.org/abs/1408.5768} {arXiv:1408.5768 [hep-lat]} \BibitemShut
  {NoStop}%
\bibitem [{\citenamefont {Blondel}\ \emph {et~al.}(2019)\citenamefont
  {Blondel}, \citenamefont {Gluza}, \citenamefont {Jadach}, \citenamefont
  {Janot},\ and\ \citenamefont {Riemann}}]{Blondel:2019vdq}%
  \BibitemOpen
  \bibinfo {editor} {\bibfnamefont {A.}~\bibnamefont {Blondel}}, \bibinfo
  {editor} {\bibfnamefont {J.}~\bibnamefont {Gluza}}, \bibinfo {editor}
  {\bibfnamefont {S.}~\bibnamefont {Jadach}}, \bibinfo {editor} {\bibfnamefont
  {P.}~\bibnamefont {Janot}},\ and\ \bibinfo {editor} {\bibfnamefont
  {T.}~\bibnamefont {Riemann}},\ eds.,\ \href
  {https://doi.org/10.23731/CYRM-2020-003} {\emph {\bibinfo {title} {{Theory
  for the FCC-ee}: {Report on the 11th FCC-ee Workshop Theory and
  Experiments}}}},\ \bibinfo {series} {CERN Yellow Reports: Monographs}, Vol.\
  \bibinfo {volume} {3/2020}\ (\bibinfo  {publisher} {CERN},\ \bibinfo
  {address} {Geneva},\ \bibinfo {year} {2019})\ \Eprint
  {https://arxiv.org/abs/1905.05078} {arXiv:1905.05078 [hep-ph]} \BibitemShut
  {NoStop}%
\bibitem [{Tev(2016)}]{TevatronElectroweakWorkingGroup:2016lid}%
  \BibitemOpen
  \href@noop {} {\bibinfo {title} {{Combination of CDF and D0 results on the
  mass of the top quark using up $9.7\:{\rm fb}^{-1}$ at the Tevatron}}}
  (\bibinfo {year} {2016}),\ \Eprint {https://arxiv.org/abs/1608.01881}
  {arXiv:1608.01881 [hep-ex]} \BibitemShut {NoStop}%
\bibitem [{\citenamefont {Sirunyan}\ \emph
  {et~al.}(2017{\natexlab{a}})\citenamefont {Sirunyan} \emph
  {et~al.}}]{Sirunyan:2017huu}%
  \BibitemOpen
  \bibfield  {author} {\bibinfo {author} {\bibfnamefont {A.~M.}\ \bibnamefont
  {Sirunyan}} \emph {et~al.} (\bibinfo {collaboration} {CMS}),\ }\bibfield
  {title} {\bibinfo {title} {{Measurement of the top quark mass using single
  top quark events in proton-proton collisions at $\sqrt{s}= 8$ TeV}},\ }\href
  {https://doi.org/10.1140/epjc/s10052-017-4912-8} {\bibfield  {journal}
  {\bibinfo  {journal} {Eur. Phys. J. C}\ }\textbf {\bibinfo {volume} {77}},\
  \bibinfo {pages} {354} (\bibinfo {year} {2017}{\natexlab{a}})},\ \Eprint
  {https://arxiv.org/abs/1703.02530} {arXiv:1703.02530 [hep-ex]} \BibitemShut
  {NoStop}%
\bibitem [{\citenamefont {Aaboud}\ \emph
  {et~al.}(2018{\natexlab{b}})\citenamefont {Aaboud} \emph
  {et~al.}}]{Aaboud:2018wps}%
  \BibitemOpen
  \bibfield  {author} {\bibinfo {author} {\bibfnamefont {M.}~\bibnamefont
  {Aaboud}} \emph {et~al.} (\bibinfo {collaboration} {ATLAS}),\ }\bibfield
  {title} {\bibinfo {title} {{Measurement of the Higgs boson mass in the
  $H\rightarrow ZZ^* \rightarrow 4\ell$ and $H \rightarrow \gamma\gamma$
  channels with $\sqrt{s}=13$ TeV $pp$ collisions using the ATLAS detector}},\
  }\href {https://doi.org/10.1016/j.physletb.2018.07.050} {\bibfield  {journal}
  {\bibinfo  {journal} {Phys. Lett. B}\ }\textbf {\bibinfo {volume} {784}},\
  \bibinfo {pages} {345} (\bibinfo {year} {2018}{\natexlab{b}})},\ \Eprint
  {https://arxiv.org/abs/1806.00242} {arXiv:1806.00242 [hep-ex]} \BibitemShut
  {NoStop}%
\bibitem [{ATL(2020)}]{ATLAS:2020coj}%
  \BibitemOpen
  \href@noop {} {\bibinfo {title} {{Measurement of the Higgs boson mass in the
  $H \rightarrow ZZ^* \rightarrow 4\ell$ decay channel with $\sqrt{s}=13$ TeV
  $pp$ collisions using the ATLAS detector at the LHC}}} (\bibinfo {year}
  {2020})\BibitemShut {NoStop}%
\bibitem [{\citenamefont {Sirunyan}\ \emph
  {et~al.}(2017{\natexlab{b}})\citenamefont {Sirunyan} \emph
  {et~al.}}]{Sirunyan:2017exp}%
  \BibitemOpen
  \bibfield  {author} {\bibinfo {author} {\bibfnamefont {A.~M.}\ \bibnamefont
  {Sirunyan}} \emph {et~al.} (\bibinfo {collaboration} {CMS}),\ }\bibfield
  {title} {\bibinfo {title} {{Measurements of properties of the Higgs boson
  decaying into the four-lepton final state in pp collisions at $ \sqrt{s}=13 $
  TeV}},\ }\href {https://doi.org/10.1007/JHEP11(2017)047} {\bibfield
  {journal} {\bibinfo  {journal} {JHEP}\ }\textbf {\bibinfo {volume} {11}},\
  \bibinfo {pages} {047}},\ \Eprint {https://arxiv.org/abs/1706.09936}
  {arXiv:1706.09936 [hep-ex]} \BibitemShut {NoStop}%
\bibitem [{\citenamefont {Sirunyan}\ \emph {et~al.}(2020)\citenamefont
  {Sirunyan} \emph {et~al.}}]{Sirunyan:2020xwk}%
  \BibitemOpen
  \bibfield  {author} {\bibinfo {author} {\bibfnamefont {A.~M.}\ \bibnamefont
  {Sirunyan}} \emph {et~al.} (\bibinfo {collaboration} {CMS}),\ }\bibfield
  {title} {\bibinfo {title} {{A measurement of the Higgs boson mass in the
  diphoton decay channel}},\ }\href
  {https://doi.org/10.1016/j.physletb.2020.135425} {\bibfield  {journal}
  {\bibinfo  {journal} {Phys. Lett. B}\ }\textbf {\bibinfo {volume} {805}},\
  \bibinfo {pages} {135425} (\bibinfo {year} {2020})},\ \Eprint
  {https://arxiv.org/abs/2002.06398} {arXiv:2002.06398 [hep-ex]} \BibitemShut
  {NoStop}%
\bibitem [{\citenamefont {Aaij}\ \emph {et~al.}(2021)\citenamefont {Aaij} \emph
  {et~al.}}]{LHCb:2021bjt}%
  \BibitemOpen
  \bibfield  {author} {\bibinfo {author} {\bibfnamefont {R.}~\bibnamefont
  {Aaij}} \emph {et~al.} (\bibinfo {collaboration} {LHCb}),\ }\href@noop {}
  {\bibinfo {title} {{Measurement of the $W$ boson mass}}} (\bibinfo {year}
  {2021}),\ \Eprint {https://arxiv.org/abs/2109.01113} {arXiv:2109.01113
  [hep-ex]} \BibitemShut {NoStop}%
\bibitem [{\citenamefont {Aaltonen}\ \emph {et~al.}(2018)\citenamefont
  {Aaltonen} \emph {et~al.}}]{Aaltonen:2018dxj}%
  \BibitemOpen
  \bibfield  {author} {\bibinfo {author} {\bibfnamefont {T.~A.}\ \bibnamefont
  {Aaltonen}} \emph {et~al.} (\bibinfo {collaboration} {CDF, D0}),\ }\bibfield
  {title} {\bibinfo {title} {{Tevatron Run II combination of the effective
  leptonic electroweak mixing angle}},\ }\href
  {https://doi.org/10.1103/PhysRevD.97.112007} {\bibfield  {journal} {\bibinfo
  {journal} {Phys. Rev. D}\ }\textbf {\bibinfo {volume} {97}},\ \bibinfo
  {pages} {112007} (\bibinfo {year} {2018})},\ \Eprint
  {https://arxiv.org/abs/1801.06283} {arXiv:1801.06283 [hep-ex]} \BibitemShut
  {NoStop}%
\bibitem [{\citenamefont {Janot}\ and\ \citenamefont
  {Jadach}(2020)}]{Janot:2019oyi}%
  \BibitemOpen
  \bibfield  {author} {\bibinfo {author} {\bibfnamefont {P.}~\bibnamefont
  {Janot}}\ and\ \bibinfo {author} {\bibfnamefont {S.}~\bibnamefont {Jadach}},\
  }\bibfield  {title} {\bibinfo {title} {{Improved Bhabha cross section at LEP
  and the number of light neutrino species}},\ }\href
  {https://doi.org/10.1016/j.physletb.2020.135319} {\bibfield  {journal}
  {\bibinfo  {journal} {Phys. Lett. B}\ }\textbf {\bibinfo {volume} {803}},\
  \bibinfo {pages} {135319} (\bibinfo {year} {2020})},\ \Eprint
  {https://arxiv.org/abs/1912.02067} {arXiv:1912.02067 [hep-ph]} \BibitemShut
  {NoStop}%
\bibitem [{\citenamefont {Bernreuther}\ \emph {et~al.}(2017)\citenamefont
  {Bernreuther}, \citenamefont {Chen}, \citenamefont {Dekkers}, \citenamefont
  {Gehrmann},\ and\ \citenamefont {Heisler}}]{Bernreuther:2016ccf}%
  \BibitemOpen
  \bibfield  {author} {\bibinfo {author} {\bibfnamefont {W.}~\bibnamefont
  {Bernreuther}}, \bibinfo {author} {\bibfnamefont {L.}~\bibnamefont {Chen}},
  \bibinfo {author} {\bibfnamefont {O.}~\bibnamefont {Dekkers}}, \bibinfo
  {author} {\bibfnamefont {T.}~\bibnamefont {Gehrmann}},\ and\ \bibinfo
  {author} {\bibfnamefont {D.}~\bibnamefont {Heisler}},\ }\bibfield  {title}
  {\bibinfo {title} {{The forward-backward asymmetry for massive bottom quarks
  at the $Z$ peak at next-to-next-to-leading order QCD}},\ }\href
  {https://doi.org/10.1007/JHEP01(2017)053} {\bibfield  {journal} {\bibinfo
  {journal} {JHEP}\ }\textbf {\bibinfo {volume} {01}},\ \bibinfo {pages}
  {053}},\ \Eprint {https://arxiv.org/abs/1611.07942} {arXiv:1611.07942
  [hep-ph]} \BibitemShut {NoStop}%
\bibitem [{\citenamefont {Abe}\ \emph {et~al.}(2000)\citenamefont {Abe} \emph
  {et~al.}}]{Abe:2000uc}%
  \BibitemOpen
  \bibfield  {author} {\bibinfo {author} {\bibfnamefont {K.}~\bibnamefont
  {Abe}} \emph {et~al.} (\bibinfo {collaboration} {SLD}),\ }\bibfield  {title}
  {\bibinfo {title} {{First direct measurement of the parity violating coupling
  of the Z0 to the s quark}},\ }\href
  {https://doi.org/10.1103/PhysRevLett.85.5059} {\bibfield  {journal} {\bibinfo
   {journal} {Phys. Rev. Lett.}\ }\textbf {\bibinfo {volume} {85}},\ \bibinfo
  {pages} {5059} (\bibinfo {year} {2000})},\ \Eprint
  {https://arxiv.org/abs/hep-ex/0006019} {arXiv:hep-ex/0006019} \BibitemShut
  {NoStop}%
\bibitem [{\citenamefont {Keshavarzi}\ \emph {et~al.}(2020)\citenamefont
  {Keshavarzi}, \citenamefont {Marciano}, \citenamefont {Passera},\ and\
  \citenamefont {Sirlin}}]{Keshavarzi:2020bfy}%
  \BibitemOpen
  \bibfield  {author} {\bibinfo {author} {\bibfnamefont {A.}~\bibnamefont
  {Keshavarzi}}, \bibinfo {author} {\bibfnamefont {W.~J.}\ \bibnamefont
  {Marciano}}, \bibinfo {author} {\bibfnamefont {M.}~\bibnamefont {Passera}},\
  and\ \bibinfo {author} {\bibfnamefont {A.}~\bibnamefont {Sirlin}},\
  }\bibfield  {title} {\bibinfo {title} {{Muon $g-2$ and $\Delta \alpha$
  connection}},\ }\href {https://doi.org/10.1103/PhysRevD.102.033002}
  {\bibfield  {journal} {\bibinfo  {journal} {Phys. Rev. D}\ }\textbf {\bibinfo
  {volume} {102}},\ \bibinfo {pages} {033002} (\bibinfo {year} {2020})},\
  \Eprint {https://arxiv.org/abs/2006.12666} {arXiv:2006.12666 [hep-ph]}
  \BibitemShut {NoStop}%
\bibitem [{\citenamefont {Flacher}\ \emph {et~al.}(2009)\citenamefont
  {Flacher}, \citenamefont {Goebel}, \citenamefont {Haller}, \citenamefont
  {Hocker}, \citenamefont {Monig} \emph {et~al.}}]{Flacher:2008zq}%
  \BibitemOpen
  \bibfield  {author} {\bibinfo {author} {\bibfnamefont {H.}~\bibnamefont
  {Flacher}}, \bibinfo {author} {\bibfnamefont {M.}~\bibnamefont {Goebel}},
  \bibinfo {author} {\bibfnamefont {J.}~\bibnamefont {Haller}}, \bibinfo
  {author} {\bibfnamefont {A.}~\bibnamefont {Hocker}}, \bibinfo {author}
  {\bibfnamefont {K.}~\bibnamefont {Monig}}, \emph {et~al.},\ }\bibfield
  {title} {\bibinfo {title} {{Revisiting the global electroweak fit of the
  standard bodel and beyond with Gfitter}},\ }\href
  {https://doi.org/10.1140/epjc/s10052-009-0966-6,
  10.1140/epjc/s10052-011-1718-y} {\bibfield  {journal} {\bibinfo  {journal}
  {Eur.Phys.J.}\ }\textbf {\bibinfo {volume} {C60}},\ \bibinfo {pages} {543}
  (\bibinfo {year} {2009})},\ \Eprint {https://arxiv.org/abs/0811.0009}
  {arXiv:0811.0009 [hep-ph]} \BibitemShut {NoStop}%
\bibitem [{\citenamefont {Borsanyi}\ \emph {et~al.}(2021)\citenamefont
  {Borsanyi} \emph {et~al.}}]{Borsanyi:2020mff}%
  \BibitemOpen
  \bibfield  {author} {\bibinfo {author} {\bibfnamefont {S.}~\bibnamefont
  {Borsanyi}} \emph {et~al.},\ }\bibfield  {title} {\bibinfo {title} {{Leading
  hadronic contribution to the muon magnetic moment from lattice QCD}},\ }\href
  {https://doi.org/10.1038/s41586-021-03418-1} {\bibfield  {journal} {\bibinfo
  {journal} {Nature}\ }\textbf {\bibinfo {volume} {593}},\ \bibinfo {pages}
  {51} (\bibinfo {year} {2021})},\ \Eprint {https://arxiv.org/abs/2002.12347}
  {arXiv:2002.12347 [hep-lat]} \BibitemShut {NoStop}%
\bibitem [{\citenamefont {Buttazzo}\ \emph {et~al.}(2013)\citenamefont
  {Buttazzo}, \citenamefont {Degrassi}, \citenamefont {Giardino}, \citenamefont
  {Giudice}, \citenamefont {Sala}, \citenamefont {Salvio},\ and\ \citenamefont
  {Strumia}}]{Buttazzo:2013uya}%
  \BibitemOpen
  \bibfield  {author} {\bibinfo {author} {\bibfnamefont {D.}~\bibnamefont
  {Buttazzo}}, \bibinfo {author} {\bibfnamefont {G.}~\bibnamefont {Degrassi}},
  \bibinfo {author} {\bibfnamefont {P.~P.}\ \bibnamefont {Giardino}}, \bibinfo
  {author} {\bibfnamefont {G.~F.}\ \bibnamefont {Giudice}}, \bibinfo {author}
  {\bibfnamefont {F.}~\bibnamefont {Sala}}, \bibinfo {author} {\bibfnamefont
  {A.}~\bibnamefont {Salvio}},\ and\ \bibinfo {author} {\bibfnamefont
  {A.}~\bibnamefont {Strumia}},\ }\bibfield  {title} {\bibinfo {title}
  {{Investigating the near-criticality of the Higgs boson}},\ }\href
  {https://doi.org/10.1007/JHEP12(2013)089} {\bibfield  {journal} {\bibinfo
  {journal} {JHEP}\ }\textbf {\bibinfo {volume} {12}},\ \bibinfo {pages}
  {089}},\ \Eprint {https://arxiv.org/abs/1307.3536} {arXiv:1307.3536 [hep-ph]}
  \BibitemShut {NoStop}%
\end{thebibliography}%
 
\end{document}